\documentclass[pra,twocolumn,preprintnumbers,superscriptaddress,amsmath,amssymb]{revtex4-1}
\usepackage{graphicx}
\usepackage{subfigure}
\graphicspath{{figures/}}
\usepackage{mathrsfs}
\usepackage{amsfonts}
\usepackage{times}
\usepackage{amsmath}
\usepackage{leftidx}
\usepackage{color}
\usepackage[colorlinks,linkcolor=blue,citecolor=blue]{hyperref}

\newcommand{\ind}{\operatorname{ind}}
\newcommand{\Tr}{\operatorname{Tr}}

\usepackage{braket}
\usepackage{mathtools}
\usepackage{physics}
\usepackage[normalem]{ulem}

\definecolor{purple}{rgb}{.6,.1,.6}
\definecolor{darkgreen}{rgb}{.1,.6,.1}
\definecolor{gray}{rgb}{.5,.5,.5}

\DeclareMathOperator{\sign}{sign}
\DeclareMathOperator{\diag}{diag}
\DeclarePairedDelimiter{\mean}{\langle}{\rangle}

\begin{document}

\title{Bound states and photon emission in non-Hermitian nanophotonics}
\author{Zongping Gong}
\affiliation{Max-Planck-Institut f\"ur Quantenoptik, Hans-Kopfermann-Stra{\ss}e 1, D-85748 Garching, Germany}
\affiliation{Munich Center for Quantum Science and Technology, Schellingstra{\ss}e 4, 80799 M\"unchen, Germany}
\author{Miguel Bello}
\affiliation{Max-Planck-Institut f\"ur Quantenoptik, Hans-Kopfermann-Stra{\ss}e 1, D-85748 Garching, Germany}
\affiliation{Munich Center for Quantum Science and Technology, Schellingstra{\ss}e 4, 80799 M\"unchen, Germany}
\author{Daniel Malz}
\affiliation{Max-Planck-Institut f\"ur Quantenoptik, Hans-Kopfermann-Stra{\ss}e 1, D-85748 Garching, Germany}
\affiliation{Munich Center for Quantum Science and Technology, Schellingstra{\ss}e 4, 80799 M\"unchen, Germany}
\author{Flore K. Kunst}
\affiliation{Max-Planck-Institut f\"ur Quantenoptik, Hans-Kopfermann-Stra{\ss}e 1, D-85748 Garching, Germany}
\affiliation{Munich Center for Quantum Science and Technology, Schellingstra{\ss}e 4, 80799 M\"unchen, Germany}
\affiliation{Max Planck Institute for the Science of Light, Staudtstra\ss e 2, 91058 Erlangen, Germany}
\date{\today}

\begin{abstract}
We establish a general framework for studying the bound states and the photon-emission dynamics of quantum emitters coupled to structured nanophotonic lattices with engineered dissipation (loss). In the single-excitation sector, the system can be described exactly by a non-Hermitian formalism. We have pointed out in the accompanying letter [Gong \emph{et al}., arXiv:2205.05479] that a single emitter coupled to a one-dimensional non-Hermitian lattice may already exhibit anomalous behaviors without Hermitian counterparts. Here we provide further detail on these observations. We also present several additional examples on the cases with multiple quantum emitters or in higher dimensions. Our work unveils the tip of the iceberg of the rich non-Hermitian phenomena in dissipative nanophotonic systems.
\end{abstract}
\maketitle


\section{Introduction}
Controlling the interaction between atoms (few-level systems) and light (bosonic fields) is a central goal in quantum optics. Recently, there has been considerable interest in controlling the propagation of light, and thus its interaction with atoms, through nanophotonic structures or atomic arrays~\cite{Chang2018}. In addition, there are longstanding efforts to create synthetic quantum optical systems comprised of artificial atoms coupled to arrays of bosonic modes, for example transmon qubits coupled to superconducting circuits~\cite{You2011,Liu2017a,Sundaresan2019,Kim2021} or surface acoustic waves~\cite{Manenti2017}, or ultracold atoms in state-dependent lattices~\cite{DeVega2008,Navarrete-Benlloch2011,Krinner2018}.
Engineering the bath has the potential of dramatically improving key figures of merit, such as storage fidelity~\cite{Asenjo2017}, or to enable novel devices, for example to simulate spin systems with long-range interactions~\cite{Douglas2015} or interactions with exotic spatial profiles \cite{Bello2022}. These successes motivate the search for other design paradigms, with novel physics, that may be exploited in future devices.


In a different context, there is remarkable recent progress on understanding non-Hermitian (NH) systems \cite{Bender2007,Ramy2018,Kunst2021,Ashida2021}, which 
may exhibit peculiar properties without Hermitian counterparts \cite{Hatano1996,Bender1998,Rudner2009,Longhi2009,Hughes2011,Sato2011,Malzard2015,Tony2016,Xu2017,Leykam2017,Cerjan2018,Shen2018,Kunst2018,Yao2018,Gong2018,Lee2019,Yoshida2019,Lee2019a,Kawabata2019a,Borgnia2020}. One of the most well-known unique features is exceptional points (EPs), which are specific points in the parameter space where the NH Hamiltonian becomes non-diagonalizable and some of its eigenstates coalesce \cite{Heiss2012}. In lattice systems, the parameter space is typically the Brillouin zone and an EP may appear at a band touching point. Depending on the symmetry of the Hamiltonian and the dimension of parameter space, the EPs may form curves, surfaces or even higher-dimensional objects \cite{Mandal2021,Delplace2021,Kunst2022}. More recently, the NH skin effect \cite{Yao2018}, which refers to the phenomenon where the ``bulk" eigenstates of a NH lattice localize at the boundary under open boundary conditions (OBCs), has attracted considerable interest. In one dimension (1D), the skin effect has been found to originate from the point-gap topology and may be enriched by symmetries \cite{Gong2018,Borgnia2020,Okuma2020,Zhang2020}. In higher dimensions, the skin effect has been shown to appear as long as the energy spectrum covers areas rather than curves \cite{Zhang2022}, which is typically the case. Some unique NH phenomena have been observed in various experimental platforms, including but not restricted to mechanical \cite{Bender2013,Xu2016,Martin2019,Ghatak2020,Hu2021,Singhal2022}, optical or photonic \cite{Guo2009,Peng2014,Segev2015,Doppler2016,Weimann2017,Zhou2018,Ozawa2019,Cerjan2019,Miri2019,Ozdemir2019,Xiao2020,Fan2021}, atomic \cite{Peng2016,Li2019,Ren2022,Rosa2022}, and electric or superconducting circuit \cite{Murch2019,Helbig2020,Hofmann2020} systems.

In this work, we merge these two emergent fields by studying quantum emitters in NH baths in a relatively comprehensive manner. While there are a few previous studies about this topic 
(see, e.g., Refs.~\cite{Longhi2016,Roccati2021}), all of them focus on rather specific (classes of) models. Moreover, the role of genuine NH topology \cite{Gong2018,Lee2019a,Kawabata2019a} remains fully unexplored. Here, we establish a general framework encompassing all the related previous works. We restrict ourselves to the single-excitation sector, for which the NH description becomes exact. We focus not only on the bound states \cite{Tao2016} but also on the photon and emitter dynamics \cite{Alejandro2017a,Alejandro2017b}, keeping in mind a special concern on genuine NH phenomena such as point-gap band topology. While it is beyond the scope of the present work, our framework also applies to the many-body regime \cite{Tao2016,Asenjo2017,Bello2019,Mahmoodian2020}, which could be a fruitful topic for future studies. 

We provide a few case studies on several minimal NH models, finding that they already exhibit unexpectedly rich phenomena. As already highlighted in Ref.~\cite{Gong2022}, the NH skin effect results in the so-called ``hidden" bound states that behave very differently from the usual bound states in Hermitian systems. For example, they have the remarkable property that their energy is exactly pinned to the emitter detuning. This property turns out to remain valid for multiple emitters. In addition, the dynamics of a photon emitted into such a NH bath may differ a lot from free propagation, as can be understood from the generalized Brillouin zone (GBZ), a concept originally developed to explain the skin effect quantitatively~\cite{Yao2017,Yokomizo2019}. Also, we show how to utilize the NH degrees of freedom to realize non-exponential (typically algebraic) emitter decay, which is usually invisible in (one-dimensional) Hermitian systems due to the existence of stable bound states. In particular, we provide further detail on the passive $PT$-symmetric lattice studied in Ref.~\cite{Gong2022} and quantitatively explain the exponents of algebraic decay. We also provide a few more examples exhibiting non-exponential decay in both 1D and 2D. 

The rest of the paper is structured as follows. In Sec.~\ref{Sec:GF}, we introduce the general framework for describing two-level emitters coupled to a NH bath with both coherent hopping and collective one-body loss. We also present a formal solution within the single-excitation sector. In Sec.~\ref{Sec:CSI}, we present the first case study on the general Hatano-Nelson model, extending the unidirectional limit considered in Ref.~\cite{Gong2022}. This example showcases the impact of NH topology. In Sec.~\ref{Sec:CSII}, we present the second case study on a Wick-rotated (i.e., multiplied by $i$) Hermitian 1D model rendering it fully anti-Hermitian. We also discuss some general features of such a construction based on Wick-rotation. In Sec.~\ref{Sec:CSIII}, we present the third case study on two models with EPs in 1D and 2D. In particular, we provide quantitative explanations to the algebraic atom decays observed in Ref.~\cite{Gong2022}. In Sec.~\ref{Sec:Dis}, we discuss some common features of these models as well as the possible experimental implementations. Finally, we summarize the paper and provide an outlook in Sec.~\ref{Sec:CO}.

\section{General framework}
\label{Sec:GF}
In this section, we present a general framework for studying quantum emitters in nanophotonic lattices that are completely captured by a NH description. We also provide a formal analytic solution to the eigenstate problem as well as the nonequilibrium dynamics.

\subsection{Equation of motion}
\begin{figure}
    \centering
    \includegraphics[width=6cm, clip]{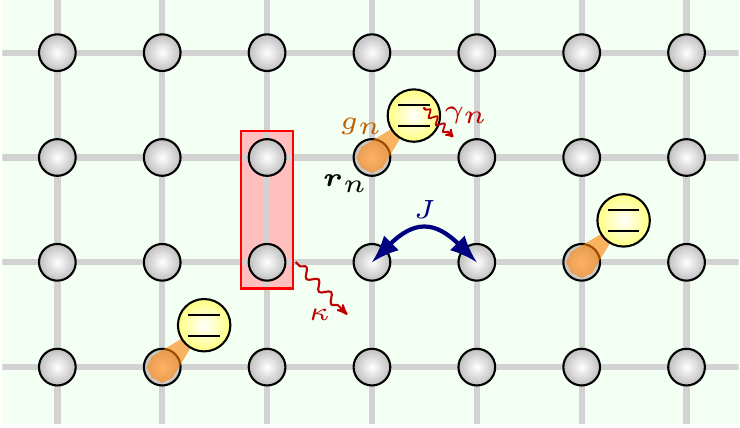}
    \caption{In our general setup, two-level atoms are coupled to a NH lattice with both coherent hopping $J$ (blue arrow) and collective loss $\kappa$ (red box), which are described by Eqs.~(\ref{Hp}) and (\ref{eq:dissipator_photon_loss}), respectively. The atoms themselves undergo spontaneous decay, as is captured by Eq.~(\ref{ad}). While not shown in the figure, the atom-photon detuning is generally nonzero [cf. Eq.~(\ref{Ha})].
    }
    \label{fig:setup}
\end{figure}

We consider $N$ (not necessarily identical) quantum emitters, modeled as two-level atoms, coupled to a $d$-dimensional nanophotonic lattice $\Lambda\subset\mathbb{Z}^d$ with engineered dissipation. Under the Markovian and rotating-wave approximations, in the rotating frame, the entire equation of motion is given by the Lindblad master equation
\begin{equation}
\dot{\hat\rho}_t=-i[\hat H_{\rm a}+\hat H_{\rm p}+\hat V,\hat\rho_t]+\mathcal{L}_{\rm a}\hat\rho_t+\mathcal{L}_{\rm p}\hat\rho_t.
\label{ME}
\end{equation}
Here the coherent part consists of three terms. The atom Hamiltonian reads
\begin{equation}
\hat H_{\rm a}=\sum^N_{n=1}\Delta^{\rm c}_n  \hat\sigma^{ee}_n,
\label{Ha}
\end{equation}
where $\hat\sigma^{ww'}_n\equiv|w_n\rangle\langle w'_n|$ ($w,w'=e,g$) and $\Delta^{\rm c}_n$ is the coherent detuning of the $n$th atom. The photon Hamiltonian takes a general (number-conserving) quadratic form
\begin{equation}
\hat H_{\rm p}=
\sum_{\boldsymbol{r}, \boldsymbol{r}'\in\Lambda} \sum_{s,s'\in I} J_{\boldsymbol{r}s,\boldsymbol{r}'s'} \hat a^\dag_{\boldsymbol{r}s} \hat a_{\boldsymbol{r}'s'},
\label{Hp}
\end{equation}
where $I$ is a set of internal states per unit cell that may involve, e.g., polarization or/and sublattice degrees of freedom, and $\hat a_{\boldsymbol{r}s}$ annihilates a photon with internal state $s$ at $\boldsymbol{r}$. The photon-atom interaction is assumed to be local (on-site): 
\begin{equation}
\hat V=\sum^N_{n=1}\sum_{s\in I} (g_{ns}\hat\sigma^{ge}_n \hat a^\dag_{\boldsymbol{r}_n s} + {\rm H.c.}),
\label{Vpa}
\end{equation}
where $\boldsymbol{r}_n$ labels the unit cell in which the $n$th atom is located and $g_{ns}$ is the coupling strength (also known as the single-photon Rabi frequency) between the $n$th atom and photon with internal state $s$. Note that Eq.~(\ref{Vpa}) conserves the total number of atom and photon excitations. The dissipative part consists of individual atom decays:
\begin{equation}
\mathcal{L}_{\rm a}=\sum^N_{n=1}\gamma_n\mathcal{D}[\hat\sigma^{ge}_n],
\label{ad}
\end{equation}
where $\gamma_n$ is the decay rate of the $n$th atom and $\mathcal{D}[\hat L]\equiv \hat L\cdot \hat L^\dag-\{\hat L^\dag \hat L,\cdot\}/2$ is the Lindblad superoperator, and collective photon loss:
\begin{equation}
\mathcal{L}_{\rm p}=\kappa\sum_{\boldsymbol{r}\in\Lambda}\sum^m_{\mu=1}\mathcal{D}[\hat L_{\boldsymbol{r}}^{\mu}],
\label{eq:dissipator_photon_loss}
\end{equation}
where $\kappa$ controls the loss rate, $\mu$ labels the dissipation channel, and $\hat L^\mu_{\boldsymbol{r}}$ takes the following form:
\begin{equation}
\hat L^{\mu}_{\boldsymbol{r}}=\sum_{\boldsymbol{r}'\in\Lambda}\sum_{s'\in I} l^{\mu}_{\boldsymbol{r},\boldsymbol{r}'s'}\hat a_{\boldsymbol{r}'s'}.
\label{Lmr}
\end{equation}
See Fig.~\ref{fig:setup} for a schematic illustration of this setup.

The effective NH Hamiltonian of Eq.~(\ref{ME}) can be computed via
\begin{align}
\begin{split}
\hat H_{\rm eff}&= \hat H_{\rm a}+\hat H_{\rm p}+\hat V \\
&- \frac{i}{2} \sum^N_{n=1}  \gamma_n \hat\sigma^{eg}_n \hat\sigma^{ge}_n- \frac{i}{2}\kappa \sum_{{\boldsymbol r} \in \Lambda} \sum^m_{\mu=1} \left(\hat L^\mu_{\bf r}\right)^\dagger \hat L^\mu_{\bf r},
\end{split}
\end{align}
and reads
\begin{equation}
\begin{split}
\hat H_{\rm eff}&= \sum^N_{n=1}\Delta_n \hat\sigma^{ee}_n + \sum_{\boldsymbol{r}, \boldsymbol{r}'\in\Lambda} \sum_{s,s'\in I} \tilde J_{\boldsymbol{r}s,\boldsymbol{r}'s'} \hat a^\dag_{\boldsymbol{r}s} \hat a_{\boldsymbol{r}'s'} \\
&+\sum^N_{n=1}\sum_{s\in I} (g_{ns}\hat\sigma^{ge}_n \hat a^\dag_{\boldsymbol{r}_n s} + {\rm H.c.}),
\end{split}
\label{Heff}
\end{equation}
where both the detuning 
\begin{equation}
\Delta_n=\Delta^{\rm c}_n - \frac{i}{2}\gamma_n
\end{equation}
and the photon hopping amplitudes
\begin{equation}
\tilde J_{\boldsymbol{r}s,\boldsymbol{r}'s'}=J_{\boldsymbol{r}s,\boldsymbol{r}'s'}- \frac{i}{2}\kappa \sum^m_{\mu=1} \sum_{\boldsymbol{r}''\in\Lambda}l^{\mu *}_{\boldsymbol{r}'',\boldsymbol{r}s} l^{\mu}_{\boldsymbol{r}'',\boldsymbol{r}'s'} \label{eq:photon_hopping_amplitudes_real_space}
\end{equation}
become complex. One can check that, since $\hat L^\mu_{\boldsymbol{r}}$'s only entail annihilation operators, starting from a single excitation state like $\rho_0=|\psi_0\rangle\langle\psi_0|$ with $|\psi_0\rangle=\hat\sigma^{eg}_1|\boldsymbol{g}\rangle\otimes|{\rm vac}\rangle$, where $|\boldsymbol{g}\rangle\equiv|g_1g_2...g_N\rangle$ and $|{\rm vac}\rangle$ is the photon vacuum, the solution to the master equation (\ref{ME}) is given by \cite{Gong2017,Alejandro2017b}
\begin{equation}
\begin{split}
\hat\rho_t&= e^{-i\hat H_{\rm eff} t}\hat\rho_0e^{iH^\dag_{\rm eff} t}+ p_t|\boldsymbol{g}\rangle\langle \boldsymbol{g}|\otimes|{\rm vac}\rangle\langle {\rm vac}|,\\
p_t&=1-\Tr[e^{-i\hat H_{\rm eff} t}\hat\rho_0e^{iH^\dag_{\rm eff} t}].
\end{split}
\end{equation}
It is thus sufficient to analyze the NH effective Hamiltonian (\ref{Heff}) if we focus on the single-excitation sector. In particular, we would like to consider the bound state
\begin{equation}
|\psi_{\rm b}\rangle=\left( \sum^N_{n=1}c^e_n \hat\sigma^{eg}_n + \sum_{\boldsymbol{r}\in\Lambda}\sum_{s\in I}c_{\boldsymbol{r}s} \hat a_{\boldsymbol{r}s}^\dag\right)|\boldsymbol{g}\rangle\otimes|{\rm vac}\rangle
\end{equation}
that satisfies $\hat H_{\rm eff}|\psi_{\rm b}\rangle=E|\psi_{\rm b}\rangle$ with nonzero atom weights and localized photon profiles near the atoms. We would also like to know the non-unitary real-time dynamics
\begin{equation}
|\psi_t\rangle=\left[ \sum^N_{n=1}c^e_n(t) \hat\sigma^{eg}_n +\sum_{\boldsymbol{r}\in\Lambda}\sum_{s\in I} c_{\boldsymbol{r}s}(t) \hat a_{\boldsymbol{r}s}^\dag\right]|\boldsymbol{g}\rangle\otimes|{\rm vac}\rangle
\end{equation}
governed by $|\psi_t\rangle=e^{-i\hat H_{\rm eff}t}|\psi_0\rangle$ starting from a localized or collective atomic excitation.

\subsection{Formal analytic solution}
Assuming that the dissipative nanophotonic lattice is translation invariant and has periodic boundary conditions (PBC), i.e., $J_{\boldsymbol{r}s,\boldsymbol{r}'s'}=J_{\boldsymbol{r}-\boldsymbol{r}',ss'}$ and $l^\mu_{\boldsymbol{r},\boldsymbol{r}'s'}=l^\mu_{\boldsymbol{r}-\boldsymbol{r}',s'}$, we can rewrite Eq.~(\ref{Heff}) into
\begin{equation}
\begin{split}
\hat H_{\rm eff}&=\sum^N_{n=1} \Delta_n\hat\sigma^{ee}_n + \sum_{\boldsymbol{k}\in{\rm B.Z.}} \hat{\boldsymbol{a}}_{\boldsymbol{k}}^\dag h_{\boldsymbol{k}}\hat{\boldsymbol{a}}_{\boldsymbol{k}}  \\
&+\frac{1}{\sqrt{|\Lambda|}}\sum^N_{n=1}\sum_{\boldsymbol{k}\in{\rm B.Z.}} (\hat\sigma^{ge}_n\hat{\boldsymbol{a}}_{\boldsymbol{k}}^\dag \boldsymbol{g}_{\boldsymbol{k}n}+{\rm H.c.}),
\end{split} \label{eq:heff_momentum}
\end{equation}
where $\hat{\boldsymbol{a}}_{\boldsymbol{k}}\equiv[\hat a_{\boldsymbol{k}s}]^{\rm T}_{s\in I}$ with $\hat a_{\boldsymbol{k}s}\equiv|\Lambda|^{-1/2}\sum_{\boldsymbol{r}\in\Lambda} e^{-i\boldsymbol{k}\cdot\boldsymbol{r}}\hat a_{\boldsymbol{r}s}$ ($|\Lambda|$: volume of $\Lambda$), $\boldsymbol{g}_{\boldsymbol{k}n}=[g_{ns} e^{-i\boldsymbol{k}\cdot\boldsymbol{r}_n}]^{\rm T}_{s\in I}$ and the NH Bloch Hamiltonian reads
\begin{equation}
[h_{\boldsymbol{k}}]_{ss'}= J_{\boldsymbol{k},ss'} -\frac{i}{2}\kappa\sum^m_{\mu=1} l^{\mu*}_{\boldsymbol{k}s}l^{\mu}_{\boldsymbol{k}s'}.
\end{equation}
Here $J_{\boldsymbol{k},ss'}=\sum_{\boldsymbol{r}\in\Lambda}J_{\boldsymbol{r}, ss'} e^{-i\boldsymbol{k}\cdot \boldsymbol{r}}$ and $l^{\mu}_{\boldsymbol{k}s}=\sum_{\boldsymbol{r}\in\Lambda} l^\mu_{\boldsymbol{r},s}e^{-i\boldsymbol{k}\cdot \boldsymbol{r}}$ 
are the Fourier transformations of the photon hopping amplitudes (\ref{Hp}) and the coefficients in the photon loss dissipators (\ref{Lmr}), respectively.

In terms of $\hat{\boldsymbol{a}}_{\boldsymbol{k}}$'s, the bound state can be rewritten as
\begin{equation}
|\psi_{\rm b}\rangle=\left( \sum^N_{n=1}c^e_n \hat\sigma^{eg}_n + \frac{1}{\sqrt{|\Lambda|}}\sum_{\boldsymbol{k}\in{\rm B.Z.}}\boldsymbol{c}_{\boldsymbol{k}}\hat{\boldsymbol{a}}_{\boldsymbol{k}}^\dag \right)|\boldsymbol{g}\rangle\otimes|{\rm vac}\rangle, \label{eq:boundstate_momentum_space}
\end{equation}
where $\boldsymbol{c}_{\boldsymbol{k}}\equiv[c_{\boldsymbol{k}s}]^{\rm T}_{s\in I}$ contains the coefficients of all the photon modes with quasi-momentum $\boldsymbol{k}$. Further introducing the detuning matrix $\Delta\equiv{\rm diag}[\Delta_1,\Delta_2,...,\Delta_N]$, the atom-coefficient vector $\boldsymbol{c}_e\equiv[c^e_1,c^e_2,...,c^e_N]^{\rm T}$ and the $|I|\times N$ coupling matrix $g_{\boldsymbol{k}}\equiv[\boldsymbol{g}_{\boldsymbol{k}1},\boldsymbol{g}_{\boldsymbol{k}2},...,\boldsymbol{g}_{\boldsymbol{k}N}]$, we can explicitly write $\hat H_{\rm eff}|\psi_{\rm b}\rangle=E|\psi_{\rm b}\rangle$ as
\begin{equation} \label{eq:solutions_eigenvalue_eqn}
\begin{split}
&\Delta \boldsymbol{c}_e + \frac{1}{|\Lambda|}\sum_{\boldsymbol{k}\in{\rm B.Z.}}g^\dag_{\boldsymbol{k}}  \boldsymbol{c}_{\boldsymbol{k}}=E \boldsymbol{c}_e,\\
&h_{\boldsymbol{k}}\boldsymbol{c}_{\boldsymbol{k}} + g_{\boldsymbol{k}} \boldsymbol{c}_e = E\boldsymbol{c}_{\boldsymbol{k}},\;\;\forall\boldsymbol{k}\in{\rm B.Z.}.
\end{split}
\end{equation}
Eliminating $\boldsymbol{c}_{\boldsymbol{k}}$'s yields
\begin{equation}
[E-\Delta -\Sigma(E) ]\boldsymbol{c}_e=0,
\end{equation}
where the self-energy $\Sigma(z)$ is an $N\times N$ matrix given by
\begin{equation}
\Sigma(z)=\frac{1}{|\Lambda|}\sum_{\boldsymbol{k}\in{\rm B.Z.}}g^\dag_{\boldsymbol{k}}(z - h_{\boldsymbol{k}})^{-1} g_{\boldsymbol{k}}. 
\label{Szf}
\end{equation}
If we take the thermodynamic limit $\Lambda\to\mathbb{Z}^d$, Eq.~(\ref{Szf}) becomes
\begin{equation}
\Sigma(z)=\int_{\rm B.Z.}\frac{d^d\boldsymbol{k}}{(2\pi)^d} g^\dag_{\boldsymbol{k}}(z - h_{\boldsymbol{k}})^{-1} g_{\boldsymbol{k}}.
\label{Sz}
\end{equation}
Since a bound state necessarily requires $\boldsymbol{c}_e\neq\boldsymbol{0}$, we have
\begin{equation}
\det[E - \Delta - \Sigma(E)]=0.
\label{det0}
\end{equation}
Given a solution $E$ to Eq.~(\ref{det0}), we can in turn determine $\boldsymbol{c}_e$ (up to normalization) and then $\boldsymbol{c}_{\boldsymbol{k}}$'s via
\begin{equation}
\boldsymbol{c}_{\boldsymbol{k}}=(E-h_{\boldsymbol{k}})^{-1}g_{\boldsymbol{k}}\boldsymbol{c}_e.
\label{ckce}
\end{equation}
The normalization of the bound state implies
\begin{equation}
    \boldsymbol{c}_e^\dag\left\{1+\frac{1}{|\Lambda|}\sum_{\boldsymbol{k}\in{\rm B.Z.}}g_{\boldsymbol{k}}^\dag [(E - h_{\boldsymbol{k}})(E^* - h^\dag_{\boldsymbol{k}})]^{-1}
    g_{\boldsymbol{k}}\right\}\boldsymbol{c}_e = 1. \label{eq:normalization}
\end{equation}
The real-space photon profile can thus be determined by the inverse Fourier transformation, i.e.,
\begin{equation}
c_{\boldsymbol{r}s}=\frac{1}{|\Lambda|}\sum_{\boldsymbol{k}\in{\rm B.Z.}} c_{\boldsymbol{k}s}e^{i\boldsymbol{k}\cdot\boldsymbol{r}}  
\xrightarrow{\Lambda\to\mathbb{Z}^d}\int_{\rm B.Z.} \frac{d^d\boldsymbol{k}}{(2\pi)^d}c_{\boldsymbol{k}s}e^{i\boldsymbol{k}\cdot\boldsymbol{r}}.
\label{crs}
\end{equation}
Provided that $E$ is not located on 
the complex photon dispersions, we can prove that (see Appendix~\ref{App:loc}) the photon profiles are exponentially localized near $\boldsymbol{r}_n$'s and thus $|\psi_{\rm b}\rangle$ is indeed a bound state.
 
To calculate the real-time dynamics, we employ the resolvent method (see Appendix~\ref{App:res}) to express $\boldsymbol{c}_e(t)\equiv[c^e_1(t),c^e_2(t),...,c^e_N(t)]^{\rm T}$ as
\begin{equation}
\boldsymbol{c}_e(t)=\frac{i}{2\pi}\int^\infty_{-\infty} dE \, G_e(E+i0^+)e^{-iEt}\boldsymbol{c}_e(0), \label{eq:cet}
\end{equation}
where the atom (emitter) Green's function $G_e(z)$ is given by
\begin{equation}
G_e(z) = \frac{1}{z - \Delta - \Sigma(z)},
\label{Ge}
\end{equation}
which is again an $N\times N$ matrix just like $\Sigma(z)$ in Eqs.~(\ref{Szf}) and (\ref{Sz}). Similarly, $\boldsymbol{c}_{\boldsymbol{k}}(t)$'s can be evaluated from
\begin{equation}
\boldsymbol{c}_{\boldsymbol{k}}(t)=\frac{i}{2\pi}\int^\infty_{-\infty} dE \, G_{\boldsymbol{k}}(E+i0^+)e^{-iEt}\boldsymbol{c}_e(0),
\label{ckt}
\end{equation}
where the photon Green's function $G_{\boldsymbol{k}}(z)$ is related to $G_e(z)$ in Eq.~(\ref{Ge}) via
\begin{equation}
G_{\boldsymbol{k}}(z)= \frac{1}{z-h_{\boldsymbol{k}}} g_{\boldsymbol{k}} G_e(z).
\label{Gk}
\end{equation}
Again, the real-space photon dynamics can be obtained by Fourier transforming $\boldsymbol{c}_{\boldsymbol{k}}(t)$ following Eq.~(\ref{crs}). Introducing
\begin{equation}
\boldsymbol{\phi}_{\boldsymbol{r}}(t)\equiv \frac{i}{2\pi}\int^\infty_{-\infty}dE\int_{\rm B.Z.}\frac{d^d\boldsymbol{k}}{(2\pi)^d}\frac{e^{i\boldsymbol{k}\cdot\boldsymbol{r}-iEt}}{E-h_{\boldsymbol{k}}+i0^+}, 
\end{equation}
which describes the bare photon propagation (in the absence of emitters) in real spacetime starting from a localized state at $\boldsymbol{r}=\boldsymbol{0}$, we can explicitly express $\boldsymbol{c}_{\boldsymbol{r}}(t)\equiv[c_{\boldsymbol{r}s}(t)]^{\rm T}_{s\in I}$ as a convolution:
\begin{equation}
\boldsymbol{c}_{\boldsymbol{r}}(t)= -i\int^t_0dt'\Phi_{\boldsymbol{r}}(t-t')\boldsymbol{c}_e(t'),
\label{crt}
\end{equation}
where the kernel function $\Phi_{\boldsymbol{r}}(t)$ is a $|I|\times N$ matrix given by
\begin{equation}
[\Phi_{\boldsymbol{r}}(t)]_{sn}= \sum_{s'\in I} [\boldsymbol{\phi}_{\boldsymbol{r}-\boldsymbol{r}_n}(t)]_{ss'} g_{n s'}. 
\end{equation}
One can check the self-consistency of Eq.~(\ref{crt}) in the short-time regime: for a sufficiently small $t$, we have $[\Phi_{\boldsymbol{r}}(t)]_{sn}\simeq \delta_{\boldsymbol{r},\boldsymbol{r}_n}g_{ns}$ and thus $c_{\boldsymbol{r}s}(t)\simeq-ig_{ns} t \sum^N_{n=1}\delta_{\boldsymbol{r},\boldsymbol{r}_n} c^e_n(0)$, exactly reproducing the leading-order approximation $\langle\boldsymbol{r}s|(-i\hat H_{\rm eff}t)|\psi_0\rangle$, where $|\boldsymbol{r}s\rangle\equiv \hat a^\dag_{\boldsymbol{r}s}|{\rm vac}\rangle$.

\section{Case study I --- non-Hermitian topology}
\label{Sec:CSI}

In our first case study, we wish to examine the effect of NH topology on the dynamics of our system by studying a model that has a nonzero spectral winding number \cite{Gong2018}. The most paradigmatic model with this feature is the Hatano-Nelson 
model \cite{Hatano1996}, which consists of a periodic 1D chain with one degree of freedom with anisotropic nearest-neighbor (NN) hoppings. To realize this model in the effective NH Hamiltonian in Eq.~(\ref{Heff}), we choose the non-local dissipator $\hat L_x = \hat a_x - i \hat a_{x+1}$ \cite{Metelmann2015, Gong2018} ($x$: unit-cell label in 1D), such that $l_{x,x'} =  \delta_{x,x'} - i \delta_{x+1,x'}$, and restrict hoppings along the 1D photonic lattice to NN only. Note that we have dropped the label $s$ because we only have one degree of freedom per site. 
We then find that the effective photon hopping amplitudes [cf. Eq.~(\ref{eq:photon_hopping_amplitudes_real_space})] read
\begin{equation}
    \tilde{J}_{x,x+1} = J - \frac{\kappa}{2}, \;\;\;\; \tilde{J}_{x,x-1} = J + \frac{\kappa}{2}, \;\;\;\; \tilde{J}_{x,x} = - i\kappa,
\end{equation}
where we set $J_{x,x+1} = J_{x-1,x} = J\in\mathbb{R}$ and $J_{x,x} = 0$, which leads to the following effective photon Hamiltonian
\begin{align}
\begin{split}
    & \hat H_{\textrm{eff},p} \equiv \sum_{x, x' \in\Lambda} \tilde J_{x, x'} \hat a^\dag_{x} \hat a_{x'} \\
    & = \sum_{x \in \Lambda} \left[\left(J - \frac{\kappa}{2}\right) \hat a^\dagger_{x} \hat a_{x+1} + \left(J + \frac{\kappa}{2}\right) \hat a^\dagger_{x+1} \hat a_{x} - i \kappa \, \hat a^\dagger_{x} \hat a_{x}\right].
    \end{split}
\end{align}
Fourier transforming this model, we find $h_k$ in Eq.~(\ref{eq:heff_momentum}) to be
\begin{equation}
\begin{split}
h_k&= \left(J-\frac{\kappa}{2}\right)e^{ik} + \left(J+\frac{\kappa}{2}\right)e^{-ik}-i \kappa \\
&= 2 J \cos k - i \kappa ( \sin k + 1),
\end{split}
\label{HNhk}
\end{equation}
which simply gives the band dispersion. The photon energy thus indeed forms a loop in complex energy plane. 
As it turns out, one can classify 
NH topological systems with point gaps using the winding number of the dispersion relation around any point $z$ in its interior \cite{Gong2018}: 
\begin{equation}
\ind(h_k - z)\equiv\int^\pi_{-\pi} \frac{dk}{2\pi i} \partial_k \ln \det (h_k - z). 
\label{ind}
\end{equation}
The Hatano-Nelson model can be in two phases: $\ind(h_k - z) = - 1$ for $J > 0$ (stronger hoppings to the right) or $\ind(h_k - z) = 1$ for $J < 0$ (stronger hoppings to the left).

In the following, we consider several emitters, each one coupled locally to a single site of the bath. Thus, the coupling function for the $n$th emitter reads $g_{k n} = g_n e^{-ik x_n}$. From Eq.~\eqref{Szf}, we can now compute the matrix elements of the self-energy as $[\Sigma(z)]_{mn} = g_m^* g_n \phi(z, x_{mn})$, where $x_{mn} \equiv x_m - x_n$ is the (signed) distance between the $m$th and $n$th emitter, and
\begin{equation}
    \phi(z, x) \equiv \frac{1}{2\pi}\int_{-\pi}^\pi dk\frac{e^{ikx}}{z-h_k} \,.
\end{equation}
The integral in this formula can be computed using residue integration (see Appendix \ref{App:zero}). 
For $\abs{J}\neq \kappa/2$, the final result can be expressed as
\begin{equation}
    \phi(z, x) = \frac{1}{\sqrt{z^2+2i\kappa z - 4J^2}}\left(y_+^{\abs{x}}\Theta_+ - y_-^{\abs{x}}\Theta_-\right) \,,
\end{equation}
where
\begin{equation}
    y_\pm = \frac{z + i\kappa \mp \sqrt{z^2+2i\kappa z - 4J^2}}{2J - \sign(x)\kappa} \,,
\end{equation}
and $\Theta_\pm \equiv \Theta(1 - \abs{y_\pm})$ ($\Theta$ denotes Heaviside's step function).
This expression is valid also for $x = 0$, which corresponds to the single-emitter case, choosing either ${\sign(0) = \pm 1}$, a result that was reported already in Ref.~\cite{Longhi2016}.
For the fully-directional Hatano-Nelson model ($\abs{J} = \kappa/2$) a similar expression can be obtained (shown in Appendix \ref{App:zero});
this specific case is studied in the companion letter \cite{Gong2022}.

As it turns out, $\phi(z, x\geq 0) = 0$ for $J>0$ [$\phi(z, x\leq 0) = 0$ for $J<0$] for $z$ inside the loop formed by the bath's dispersion relation. 
This is not a coincidence, but a general phenomenon linking the topology of the NH bath to the quantum emitter properties. 
In Appendix \ref{App:zero}, using the argument principle \cite{Ahlfors1979}, we prove that
for arbitrary 1D, single-band lattices with a finite hopping range, the self-energy vanishes in the regions of the complex plane with maximal spectral winding number.

\subsection{Bound states}

Now, we can compute the energy $E$ of the single-particle bound states  using Eq.~\eqref{det0}. For any given $E$, the photonic component (wavefunction in the bath) of the bound states, is according to Eq.~\eqref{crs}, given by $\phi(E, d)$:
\begin{equation}
    c_x = \sum^N_{n=1} c^e_n 
    g_n \phi(E, x - x_n) \,.
\end{equation}
So, in general, the photonic component of any bound state is a superposition of exponentially-localized wavefunctions around each emitter.
Importantly, $\phi(E, x)$ does not have the same decay length for the left-half ($x<0$) and right-half ($x>0$) spaces. 
In fact, as we have just mentioned, $\phi(E, x)$ may even vanish completely on one of the sides. 

\begin{figure}
    \centering
    \includegraphics[width=8.5cm, clip]{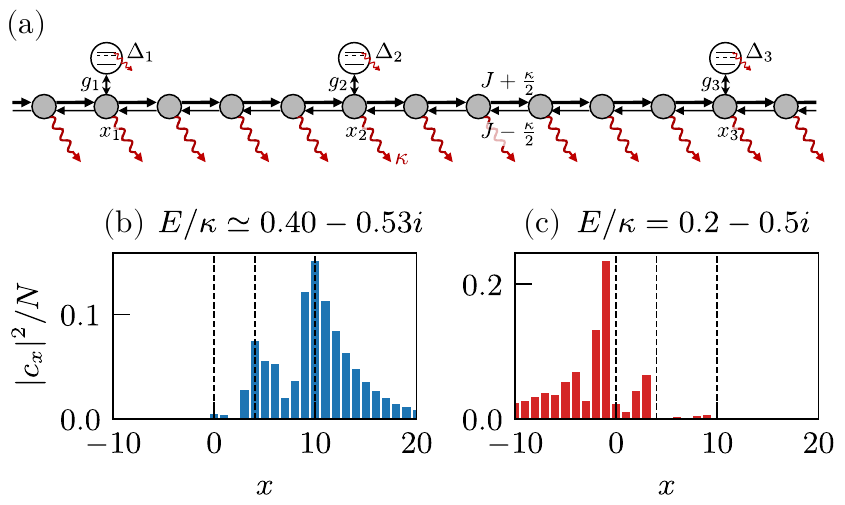}
    \caption{(a) Hatano-Nelson model with three emitters coupled at three different locations. Conventional bound state (b) and skin-like bound state (c) for such a system with parameters: $J=0.15\kappa$, $\Delta_n/\kappa = 0.05(n + 1) - 0.5i$, $n=1,2,3$,  and $g_n=0.5\kappa$. Dashed vertical lines mark the positions $x_n$ of the emitters.}
    \label{fig:HNBS}
\end{figure}

The shape of the bound states is intimately related to the characteristics of the bath and, in particular, to its topology. We can classify solutions in two classes that we call \emph{conventional bound states} if $E$ lies outside the loop and \emph{hidden bound states} if $E$ lies inside the loop. The former are composed of wavefunctions that decay on both sides of the emitters, see Fig.~\ref{fig:HNBS}(a). 

The latter are composed of fully directional wavefunctions, even if the bath has finite hopping amplitudes in both directions, and they decay towards the direction of the weaker hopping amplitudes, see Fig.~\ref{fig:HNBS}(b). They are unique to NH systems with nonzero spectral winding numbers. Strikingly, their energies are not affected by the presence of other emitters, contrary to what happens for conventional bound states (except when the bath is fully directional). This is due to the fact that $\Sigma(E)$ is strictly upper (lower) triangular if we label the emitters by increasing position for $E\in\ell$ if $J > 0$ ($J < 0$), so
\begin{equation}
    \det[E - \Delta - \Sigma(E)] = \prod^N_{n=1} (E - \Delta_n) \,, \ \text{if }E\in\ell\,.
\end{equation}
The equation $\det[E - \Delta - \Sigma(E)] = 0$ [cf. Eq.~\eqref{det0}] has, therefore, a solution $E = \Delta_n$ for every $\Delta_n \in \ell$. Here, we 
recall the possibility of having different detunings $\Delta_n$ for each emitter, and 
$\Delta\equiv\diag[\Delta_1,\Delta_2,...,\Delta_N]$.

We can recognize these hidden bound states as the NH analogues of the vacancy-like bound states discussed in the Hermitian context \cite{Leonforte2021}. Indeed, if we put a vacancy in the site to which an emitter is coupled we split the bath in two semi-infinite chains, whose eigenmodes can be computed using, e.g., the transfer-matrix method \cite{Kunst2019}. Skin modes correspond to those eigenvalues of the transfer matrix with absolute value smaller than one, which restricts their energy to the interior of $h_k$. Thus, the condition for a perfect vacancy-like bound state, namely, that the emitter detuning coincides with the energy of the vacancy mode (in this case, skin mode), is trivially satisfied whenever $\Delta$ lies inside the loop.

Interestingly, conventional bound states may require a sufficiently strong coupling constant $g$ to exist, depending on the value of $\Delta$. This is due to the fact that $[\Sigma(z)]_{mn}$ has a finite discontinuity when $z$ traverses the bath's dispersion relation. This behavior is very different from the one expected for 1D Hermitian systems, where the divergence of the real part of the self-energy at the band edges of the bath's spectrum guarantees the existence of bound states in every gap, regardless the value of $\Delta$ or $g$ \cite{Tao2016}.

\subsection{Photon-emission dynamics}
In the companion letter \cite{Gong2022}, we obtained the analytic expression for the time evolution of a photon emitted from the atomic excitation in the unidirectional limit. While it seems impossible to have an analytic solution in the general case,  we can readily perform numerical calculations. We demonstrate that the three different dynamical regimes still exist and can be well understood with the help of the GBZ 
\cite{Yao2018,Yokomizo2019}.

\begin{figure}
    \centering
    \includegraphics[width=8.5cm, clip]{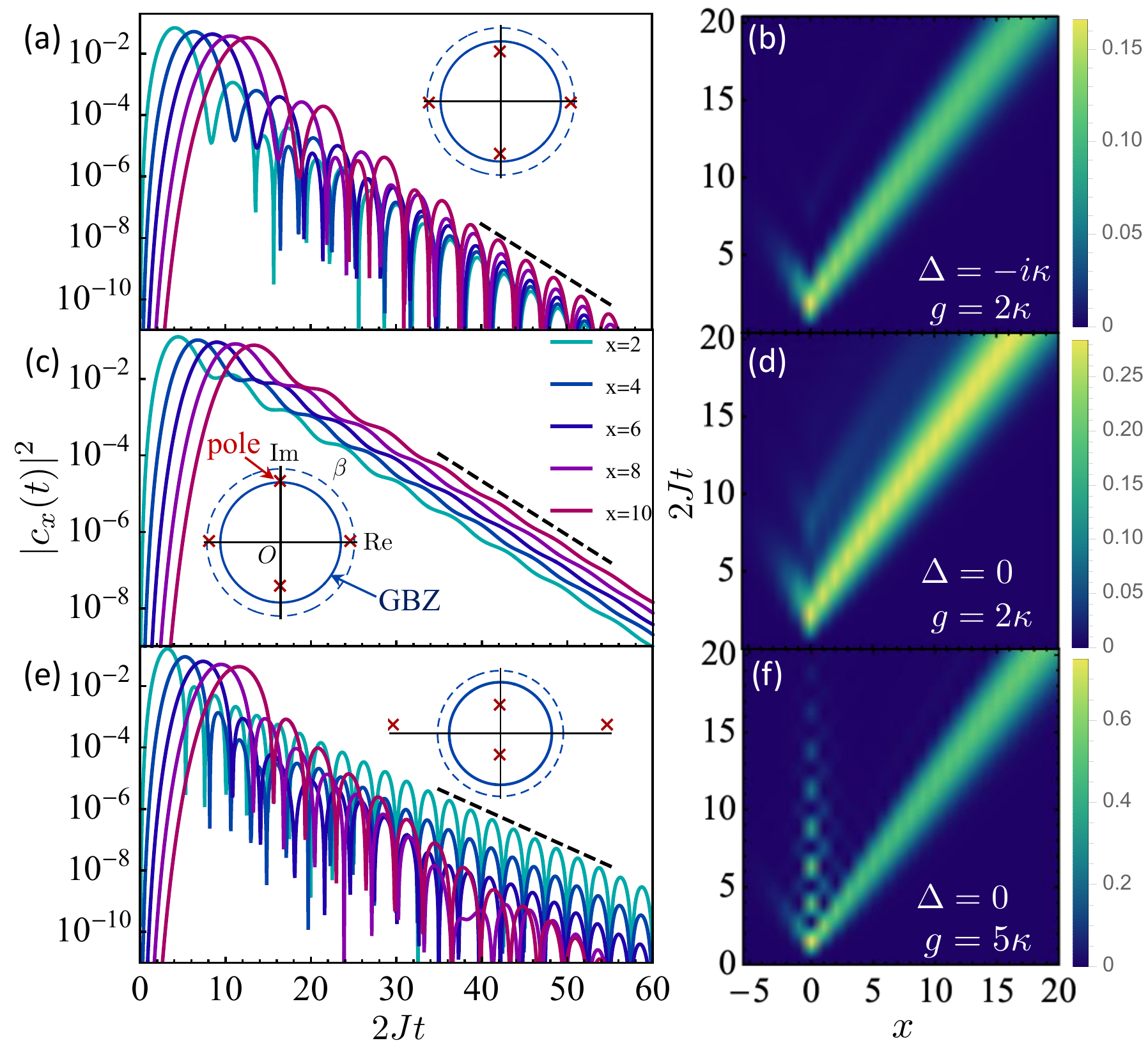}
    \caption{Real-space photon-emission dynamics for the general Hatano-Nelson model (\ref{HNhk}) with $\Delta=-i\kappa$, $g=2\kappa$ ((a) and (b)), $\Delta=0$, $g=2\kappa$ ((c) and (d)) and $\Delta=0$, $g=5\kappa$ ((e) and (f)). In all figures, $J=2.5\kappa$. Insets in (a), (c) and (e): original (dashed circle) and generalized Brillouin zone (solid circle) in the complex plane $\beta\equiv e^{-ik}$. The red crosses indicate the poles of the integrand in Eq.~(\ref{cRW}). Dashed black lines in (a), (c) and (e) correspond to the exponential decay of free propagation, the residue of the marked pole and the bound state, respectively. Just like Fig.~\ref{fig:exciting_BIC}(b), the photon profiles are rescaled by multiplying $\sqrt{2J t}$ in (b), (d) and (f) so that the probability (almost) converges in the long-time limit \cite{Gong2022}.}
    \label{fig:HNPD}
\end{figure}

As mentioned previously, a unique feature of 1D NH nanophotonic systems is that the bound states may disappear if the light-matter coupling is too weak. In addition, the hidden bound states with eigenenergies inside the loop of the bath dispersion do not contribute to the dynamics and are thus invisible. One may thus na\"ively expect that, in the absence of conventional bound states, the photon emission dynamics should resemble free propagation. This is indeed the case by choosing, e.g., $(J,\Delta, g)=(2.5,-i,2)\kappa$. The corresponding dynamics is shown in Figs.~\ref{fig:HNPD}(a) and (b). Unlike the unidirectional case, there is a nonzero leftward-propagating component, although it quickly decays. The oscillations result from the interference effect that appears already in the free-propagation dynamics, which is actually exactly solvable to be \cite{Ezawa2019}
\begin{equation}
|c^{\rm FP}_{x}(t)|^2= \left(\frac{J+\frac{\kappa}{2}}{J-\frac{\kappa}{2}}\right)^{x}{\rm J}_{x}(\sqrt{4J^2-\kappa^2}t)^2e^{-2\kappa t}
\label{HNfree}
\end{equation}
starting from $c_{x}(0)=\delta_{x0}$, i.e., a photon localized at the origin. Here ${\rm J}_{x}(z)$ is the Bessel function of the first kind and is known to exhibit sine-like oscillations for large $z$. Note that the amplification rate corresponds to the radius of the GBZ and the coefficient before $t$ is nothing but the bandwidth under the OBC.

However, if we change the detunning to be $\Delta=0$, we find a spatial amplifying dynamics with a decay rate (at a fixed site) considerably smaller than free propagation. See Figs.~\ref{fig:HNPD}(c) and (d). To gain some quantitative insights, we write down the running-wave contribution, which should dominate in the long time limit provided there is no bound state (or the bound sate has a very short lifetime):
\begin{equation}
c^{\rm RW}_{x}(t)= g\int^\pi_{-\pi}\frac{dk}{2\pi} \frac{e^{ikx - ih_k t}}{h_k-\Delta-\Sigma(h_k+i\eta_k)},
\end{equation}
where $\eta_k$ is a vanishingly small quantity such that $h_k+i\eta_k$ lies outside the loop. After straightforward calculations and the replacement $\beta\equiv e^{-ik}$, the running wave contribution can be expressed as the following contour integral:
\begin{widetext}
\begin{equation}
c^{\rm RW}_{x}(t)=\oint_{|\beta|=1} \frac{d\beta}{2\pi i\beta} \frac{ge^{-\kappa t}\beta^{-x}e^{-i[(J+\kappa/2)\beta + (J-\kappa/2)\beta^{-1}]t}[(J+\kappa/2)\beta-(J-\kappa/2)\beta^{-1}]}{(J+\kappa/2)^2\beta^2 - (J-\kappa/2)^2\beta^{-2} - (\Delta+i\kappa)[(J+\kappa/2)\beta-(J-\kappa/2)\beta^{-1}] - g^2 }.
\label{cRW}
\end{equation}
\end{widetext}
At least at large spacetime scales, one can justify from the stationary-phase approximation \cite{Bender1999} that the above contour integral at the GBZ $|\beta|=\sqrt{(J-\kappa/2)/(J+\kappa/2)}$ \cite{Kunst2021} should be like free propagation. This contribution differs from Eq.~(\ref{cRW}) only in some residues associated with the poles of the integrand sandwiched by the GBZ and the original Brillouin zone $|\beta|=1$. In the present case, there are three such poles  (cf. inset in Fig.~\ref{fig:HNPD}(c)) and the residue of the one lying on the imaginary axis turns out to be dominant and overwhelm the free-propagation-like component. We mention that there are actually two relevant poles in the previous case (cf. inset in Fig.~\ref{fig:HNPD}(a)), but their residues have the same decay rate as free propagation but smaller spatial amplification rate. Therefore, the free-propagation-like component dominates in the previous case.

If we further enhance the coupling strength to be $g=5\kappa$, we observe not only temporal but also spatial decay in the long time limit, as shown in Fig.~\ref{fig:HNPD}(e). This is simply due to the existence of slowly decaying bound states, which overwhelm the free-propagation component. Here the oscillations arise from the superposition of two bound states with different (actually opposite) real energies.

Finally, we should stress that the general Hatano-Nelson model is rather specific in the sense that its spectrum under the OBC has 
a constant 
imaginary part and is thus Hermitian-like. In a general setting, the OBC spectrum 
covers a finite range along the imaginary axis. In this case, it becomes unclear whether the GBZ is still a privileged choice, but the idea of deforming the integral contour remains applicable. We leave a complete solution to this problem for a future study.

\section{Case study II --- Wick-rotated Hermitian baths}
\label{Sec:CSII}
In our second case study, we introduce a simple recipe other than the $PT$-symmetry breaking mentioned in Ref.~\cite{Gong2022} to realize non-exponential (algebraic) emitter decay. 

\subsection{General analysis}
As pointed out in Ref.~\cite{Gong2022}, the essential ingredient needed to realize such algebraic emitter decay is a bath with a vanishing damping gap (in the following denoted as \emph{critical}), i.e., a bath whose 
spectrum touches the real axis. A simple way to design critical baths 
to consider fully \emph{anti-Hermitian} baths obtained simply by multiplying $i$ to Hermitian baths: 
\begin{equation}
\hat H_{\rm eff,p} = i \hat H_{\rm B},\;\;\;\; \hat H_{\rm B}^\dag = \hat H_{\rm B}. 
\end{equation}
From a dynamical point of view, such an operation is equivalent to making the real time imaginary, and thus may also be called a Wick rotation \cite{Ashida2021}. If $\hat H_{\rm B}$ has a negative spectrum with a band edge exactly at zero energy, then the resulting NH bath will have a vanishing damping gap. According to Eq.~(\ref{Sz}), the self-energy $\Sigma$ in the anti-Hermitian model 
is related to the self-energy for the Hermitian counterpart $\Sigma_{\rm B}$ through
\begin{equation}
\Sigma(z) = -i\Sigma_{\rm B}(-iz). 
\label{SNHH}
\end{equation}
Therefore, for real energy $E$, if 
\begin{equation}
\Sigma_{\rm B}(E \pm i 0^+) = \delta(E) \mp i\Gamma(E)/2, 
\end{equation}
with real Lamb-shift $\delta(E)$ and decay rate $\Gamma(E)$, which is always the case if the couplings $g_{\boldsymbol{k}}$ are real, we have  
\begin{equation}
\Sigma
(iE \pm 0^+) = -i\delta(E) \pm \Gamma(E)/2.
\end{equation}
Since the role of the Lamb-shift and decay rate are reversed, the decay of the quantum emitter may be dramatically changed by Wick-rotating the bath, as will be illustrated in a concrete model in the following.

\subsection{Example}
\label{Sec:Wick1D}
As a simple example, we consider a 1D NN Hermitian lattice 
described by
\begin{equation}
    \hat H_{\rm B} = -J\sum_{x\in\Lambda} \left(\hat{a}^\dag_{x + 1} \hat{a}_{x} + \hat{a}^\dag_{x} \hat{a}_{x + 1} + 2\hat{a}^\dag_{x} \hat{a}_{x}\right),
\end{equation}
whose dispersion relation is given by $h_k = -2J(\cos k + 1)$. The ``Wick-rotated'' version $\hat H_{\rm eff,p}=i\hat H_{\rm B}$ has $ih_k$ as its dispersion relation, so it is critical, with a branch point at $z_\mathrm{BP} = 0$. This model can be obtained in a 1D photonic lattice with a single site per unit cell, choosing $J_{x,x'} = 0$ and $\hat{L}_{x} = \hat{a}_{x} + \hat{a}_{x+1}$. Since the decay rate diverges as $\Gamma(i E) \propto |E|^{-1/2}$ for ${E\to 0^-}$, the algebraic decay is of the usual kind, i.e., proportional to $t^{-3}$.


\begin{figure}
	\includegraphics[width=8.5cm, clip]{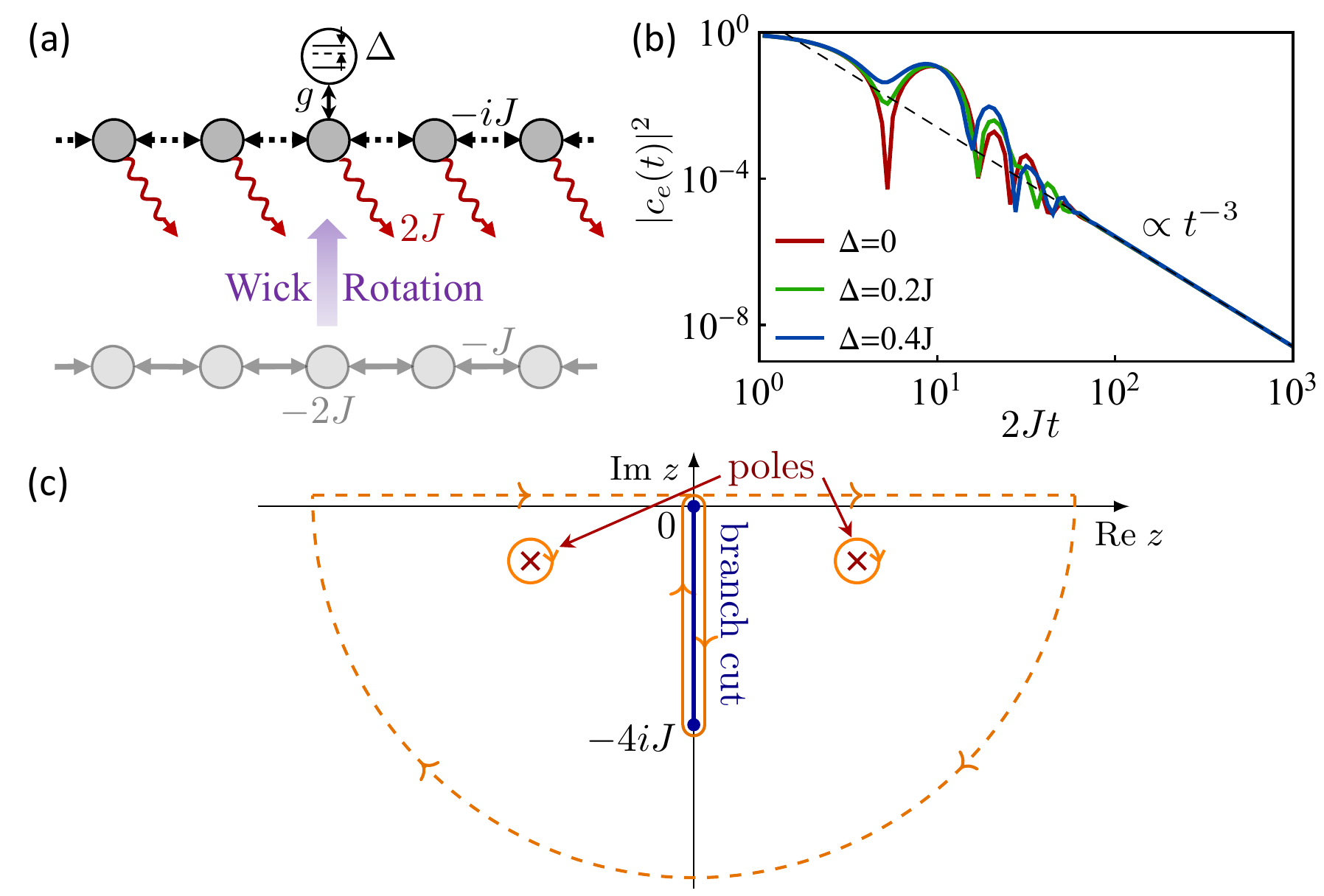}
	\caption{(a) A two-level emitter coupled to an anti-Hermitian 1D lattice obtained by Wick rotation. (b) Algebraic ($\propto t^{-3}$) emitter decay for various detunings. Here $g=J$ and the dashed line is given by Eq.~(\ref{AHt3}). (c) The original integral (\ref{eq:cet}) (dashed orange contour) for computing the emitter decay dynamics can be decomposed into the branch-cut and pole contributions (solid orange contours).}
	\label{Fig:AH1D}
\end{figure}

In fact, we can explicitly figure out the coefficient of the asymptotic algebraic decay. For a general branch cut terminating at a branch point $z_{\rm BP}$, its contribution can be calculated on the basis of the general formula \cite{Longhi2016}
\begin{equation}
c_{\rm BC}(t)\simeq \frac{e^{-i\pi (\nu+1)/2}\Gamma(\nu+1)F}{ 2\pi }
\frac{e^{-iz_{\rm BP}t}}{t^{\nu+1}},
\label{cBC}
\end{equation}
where $F$ and $\nu$ are determined by
\begin{equation}
\frac{1}{z-\Delta-\Sigma_{\rm r}(z)}-\frac{1}{z-\Delta-\Sigma_{\rm l}(z)} 
\simeq F(z-z_{\rm BP})^{\nu}
\label{Fnu}
\end{equation}
for $z$ close to $z_{\rm BP}$, with self-energy $\Sigma_{\rm r}(z)$ ($\Sigma_{\rm l}(z)$) evaluated from the right (left) side of the branch cut. For the Wick-rotated 1D lattice, according to the well-known result for 1D Hermitian NN lattice \cite{Alejandro2017b} and Eq.~(\ref{SNHH}), we can easily write down the self-energy as 
\begin{equation}
\Sigma(z)= \frac{g^2}{\sqrt{z(z+4iJ)}}, 
\end{equation}
where the square root should be taken such that $\Sigma(z)\simeq g^2/z$ for large $z$. Therefore, for $z$ near $z_{\rm BP}=0$, the left-hand side (lhs) of Eq.~(\ref{Fnu}) reads
\begin{equation}
    \frac{2g^2\sqrt{z(z+4iJ)}}{z(z+4iJ)(z-\Delta)^2-g^4}\simeq -\frac{4\sqrt{iJ}}{g^2}z^{\frac{1}{2}},
\end{equation}
implying that $\nu=1/2$ and $F=-4\sqrt{iJ}/g^2$. Substituting these results into Eq.~(\ref{cBC}), we obtain
\begin{equation}
    |c_{\rm BC}(t)|^2\simeq\frac{J}{\pi g^4}\frac{1}{t^3},
    \label{AHt3}
\end{equation}
which turns out to have no dependence on the detuning $\Delta$. As shown in Fig.~\ref{Fig:AH1D}(b), the asymptotic emitter decay is indeed given by Eq.~(\ref{AHt3}). While the exact result should involve the contributions from the poles (corresponding to bound states) and another branch cut terminating at $-4iJ$, as shown in Fig.~\ref{Fig:AH1D}(c), all of these components decay exponentially in time and are thus quickly overwhelmed by Eq.~(\ref{AHt3}).

Clearly, the long-time behavior of a quantum emitter in the Wick-rotated 1D lattice is very different from that in the original Hermitian lattice, for which $\Delta$ matters a lot and the algebraic decay is invisible \cite{Alejandro2017b}. On the other hand, it is well known in the context of NH topological phases that a Wick rotation is irrelevant, in the sense that it does not alter the topological classification at all \cite{Kawabata2019,Kawabata2019a}. There is actually no inconsistency. In the former case, we only have to focus on those eigenmodes with largest imaginary parts, which dominate the long-time dynamics. These modes may change dramatically upon the Wick rotation. In the latter case, our focus is whether two NH Hamiltonians can be continuously deformed into each other while keeping the gap open. Note that both the continuous path and gap persist upon the Wick rotation.

\section{Case study III --- exceptional points}
\label{Sec:CSIII}
In our third case study, we provide further detail on the 1D model with alternating loss mentioned in Ref.~\cite{Gong2022}. We also propose a 2D model with exceptional lines in the band structure, which turns out to exhibit both non-exponential emitter decay and diffusive photon dynamics.

\subsection{One dimension}
\begin{figure}
	\includegraphics[width=8.5cm, clip]{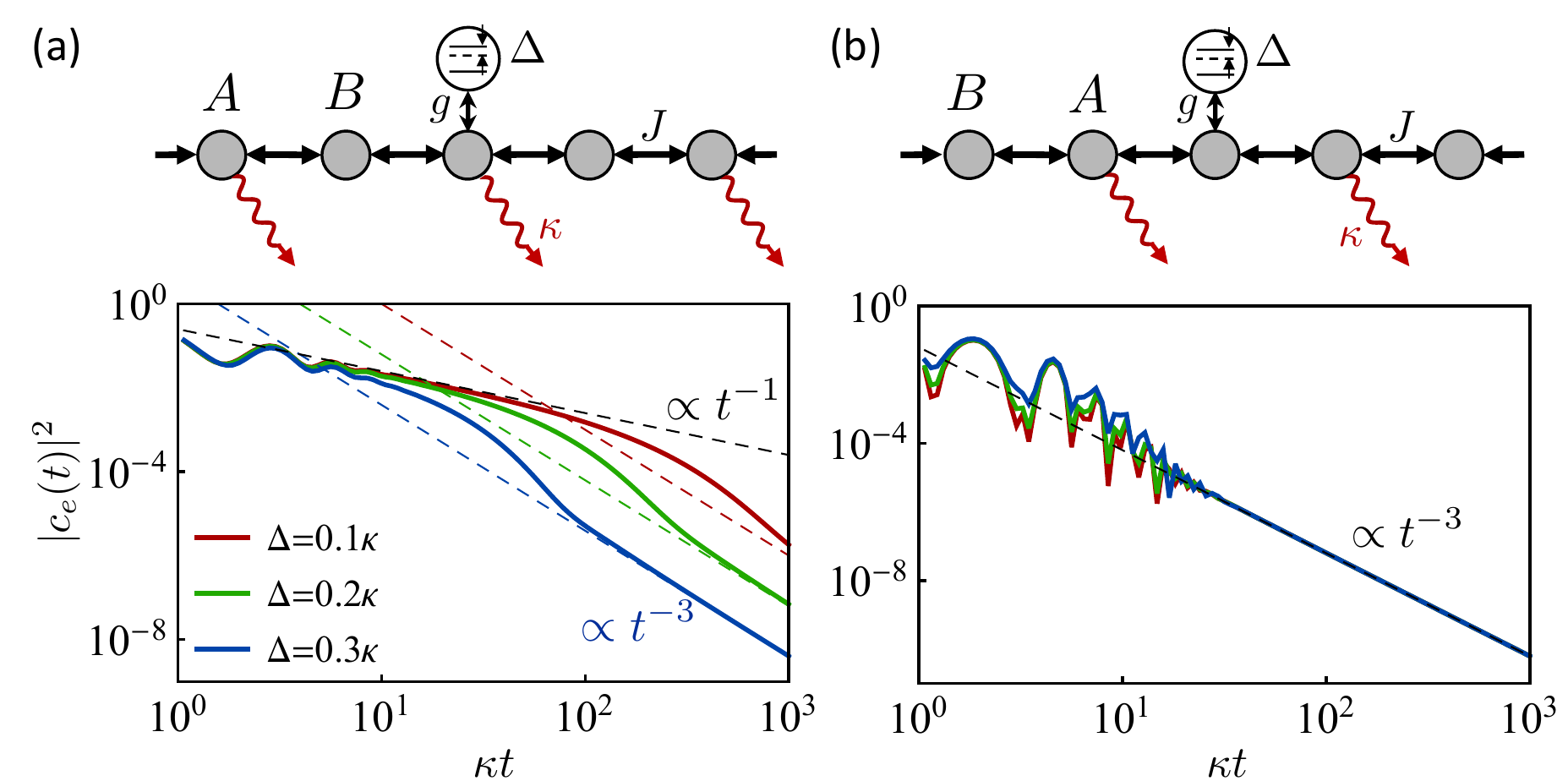}
	\caption{(a) Crossover from $t^{-1}$ to $t^{-3}$ decay for the coupling to a dissipative site under different choices of $\Delta$. The black and colored dashed lines are determined by Eqs.~(\ref{t1c}) and (\ref{t3c}), respectively. (b) Asymptotic $t^{-3}$ decay for the coupling to a non-dissipative site under different choices of $\Delta$, which only alter the short-time behaviors. The black dashed line is determined by Eq.~(\ref{ndt3c}). In both (a) and (b), $J=\kappa$ and $g=1.5\kappa$.}
	\label{Fig:crossover}
\end{figure}

We consider a 1D NH bath with a two-site unit cell. The two sublattice degrees of freedom are denoted as $A$ and $B$, as illustrated in the top of Fig.~\ref{Fig:crossover}.
Considering $\hat L_{x} = \sqrt{2} 
\hat a_{xA}$ and again restricting the hopping in the 1D photonic lattice to NN by choosing $J_{xs,x's'}=J(1-\delta_{ss'})(\delta_{xx'} + \delta_{x+1-2\delta_{s'A},x'})$,  
we find that the effective NH photon Hamiltonian reads
\begin{equation}
\hat H_{\rm eff,p}=\sum_{x\in\Lambda}[J(\hat a^\dag_{xA} + \hat a^\dag_{x+1,A})\hat a_{xB} + {\rm H.c.}-i\kappa\hat a^\dag_{xA}\hat a_{xA}].
\label{eq:alternating-diss}
\end{equation}
The corresponding NH Bloch Hamiltonian is given by
\begin{equation}
\begin{split}
h_k&=\begin{bmatrix} -i\kappa & J(1+e^{-ik}) \\ J(1+e^{ik}) & 0  \end{bmatrix}\\
&= J(1+\cos k)\hat\sigma^x+ J\sin k\hat\sigma^y-\frac{i}{2}\kappa(\hat\sigma^z+\hat\sigma^0),
\end{split}
\label{1DPT}
\end{equation}
whose energy dispersion reads
\begin{equation}
\epsilon_{k\pm} = -i\frac{\kappa}{2}\pm\sqrt{2J^2(1+\cos k) -\frac{\kappa^2}{4}}.
\end{equation}
This system exhibits a passive parity-time ($PT$) symmetry \cite{Guo2009} 
\begin{equation}
    \sigma^x \left(h_k + \frac{i}{2}\kappa\sigma_0\right)^*\sigma^x = h_k + \frac{i}{2}\kappa\sigma_0
    \label{ppt}
\end{equation}
and has two EPs at $k_{\rm EP}=\pm\arccos(\kappa^2/(8J^2)-1)$ for $J>\kappa/4$. Depending on whether the atom is located on an even (sublattice $A$) or odd (sublattice $B$) site, we have $g_A=g$, $g_B=0$ or $g_A=0$, $g_B=g$, respectively. 

For simplicity, we assume the same (real) single-photon Rabi frequency $g$ for all the emitters. Applying Eq.~(\ref{Szf}) and followed by some calculations similar to Eq.~(\ref{eq:der_self_en_hn}), we obtain the entries in the self-energy matrix to be
\begin{equation}
    \begin{split}
    \Sigma^{AA}_{mn}(z) & = \frac{g^2z}{\sqrt{\delta}}\left[y_+^{\abs{x_{mn}}}\Theta_+ - y_-^{\abs{x_{mn}}}\Theta_-\right] \,, \\
    \Sigma^{BB}_{mn}(z) & = \frac{g^2(z+i\kappa)}{\sqrt{\delta}}\left[y_+^{\abs{x_{mn}}}\Theta_+ - y_-^{\abs{x_{mn}}}\Theta_-\right] \,, \\
    \Sigma^{AB}_{mn}(z) & = \frac{g^2J}{\sqrt{\delta}}\left[F_{x_{mn}}(y_+)\Theta_+ - F_{x_{mn}}(y_-)\Theta_-\right] \,, \\
    \Sigma^{BA}_{mn}(z) & = \Sigma^{AB}_{nm}(z) \,,
    \end{split}
    \label{PTSE}
\end{equation}
where $\Theta_\pm \equiv \Theta(1-\abs{y_\pm})$, $F_x(y) \equiv y^{\abs{x}} + y^{\abs{x-1}}$, and $\delta$ is the discriminant of the second order polynomial $ay^2+by+c$ with coefficients 
\begin{equation}
a=c=-J^2,\;\;\;\;b=-2J^2 + z(z+i\kappa), 
\end{equation}
whose roots are given by $y_\pm = (-b \pm \sqrt{\delta})/(2a)$. The superscripts in $\Sigma_{mn}$ (e.g., $AA$ or $AB$) are determined by the sublattices in which the emitters are located. In this case, at any point $z$ of the complex plane, only one of the roots contributes since $y_+y_-=1$ (there are no regions with winding number other than 0). The analytic continuation of these functions to the second Riemann sheet is obtained replacing $\Theta_\pm \to \Theta_\mp$.

\subsubsection{Single-emitter dynamics}
We denote the single-emitter self-energy as $\Sigma_A(z) \equiv \Sigma^{AA}_{nn}(z)$ or $\Sigma_B(z) \equiv \Sigma^{BB}_{nn}(z)$, depending on the sublattice to which the emitter is coupled. Setting $|x_{mn}| = 0$ in the first two lines of Eq.~(\ref{PTSE}), we obtain
\begin{equation}
    \Sigma_A(z)=\frac{g^2z}{\sqrt{\delta}}(\Theta_+-\Theta_-),
    \label{SA}
\end{equation}
\begin{equation}
    \Sigma_B(z)=\frac{g^2(z+i\kappa)}{\sqrt{\delta}}(\Theta_+-\Theta_-).
    \label{SB}
\end{equation}
No matter in which sublattice the emitter is located, we expect that the long-time dynamics is dominated by the algebraic decay since there is a branch point at the origin and all the poles have negative imaginary parts. Such a situation is similar to that of the Wick-rotated 1D lattice discussed in Sec.~\ref{Sec:Wick1D} (cf. Fig.~\ref{Fig:AH1D}(c)). 

Quantitatively, we can again determine the explicit asymptotic form using Eqs.~(\ref{cBC}) and (\ref{Fnu}) (see Appendix~\ref{App:PTBC} for detail). If the emitter is in sublattice $A$ (with on-site loss) and the coupling is resonant ($\Delta=0$), by substituting Eq.~(\ref{SA}) into Eq.~(\ref{Fnu}), we obtain $\nu=-1/2$ and $|F|=4J\sqrt{\kappa}/g^2$, leading to an asymptotic $t^{-1}$ decay:
\begin{equation}
|c_{\rm BC}(t)|^2\simeq \frac{4J^2\kappa}{\pi g^4}\frac{1}{t}.
\label{t1c}
\end{equation}
Otherwise, whenever the detuning $\Delta$ is nonzero, we have $\nu=1/2$ and $|F|=g^2/(J|\Delta|^2\sqrt{\kappa})$, leading to the conventional $t^{-3}$ decay:
\begin{equation}
|c_{\rm BC}(t)|^2\simeq \frac{g^4}{16\pi |\Delta|^4 J^2\kappa} \frac{1}{t^3}.
\label{t3c}
\end{equation}
For sufficiently small $\Delta$, one may expect a crossover from $t^{-1}$ to $t^{-3}$, as indeed confirmed numerically in Fig.~\ref{Fig:crossover}(a). 

If the emitter is in sublattice $B$, we always have $\nu=1/2$ and $|F|=4J/(g^2\sqrt{\kappa})$, leading to the conventional $t^{-3}$ decay:
\begin{equation}
|c_{\rm BC}(t)|^2\simeq  \frac{J^2}{\pi g^4\kappa} \frac{1}{t^3}.
\label{ndt3c}
\end{equation}
Remarkably, similar to the case of the Wick-rotated 1D bath (\ref{AHt3}), the coefficient in Eq.~(\ref{ndt3c}) does not depend on $\Delta$, as also confirmed numerically in Fig.~\ref{Fig:crossover}(b).

\begin{figure}
	\includegraphics[width=8.5cm, clip]{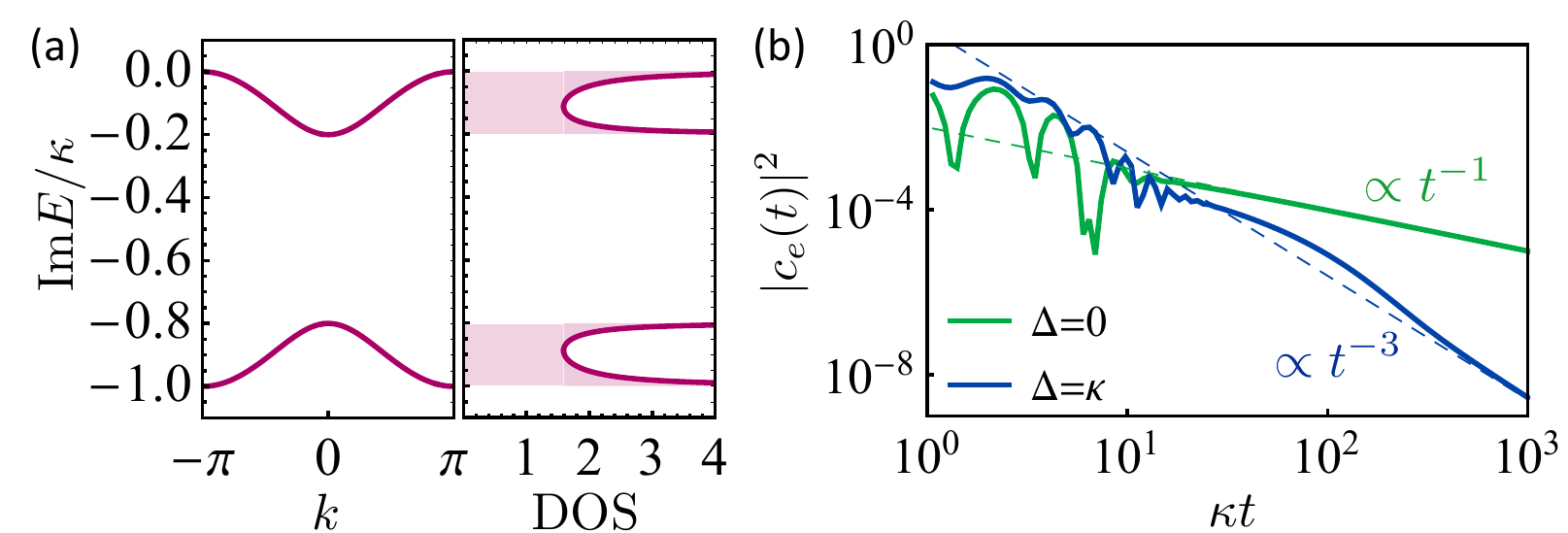}
	\caption{(a) Band dispersion and density of states (DOS) for the alternatingly lossy lattice with $J=0.2\kappa$. While the energy spectrum is purely imaginary and thus there is no EP, the DOS at zero energy (with maximal imaginary part) diverges. (b) Corresponding asymptotic $t^{-1}$ (green) and $t^{-3}$ (blue) decay for the coupling to a dissipative site with $\Delta=0$ and $\Delta=\kappa$, respectively. The green and blue dashed line are determined by Eqs.~(\ref{t1c}) and (\ref{t3c}), respectively. Here $g=1.5\kappa$.}
	\label{Fig:NoEP}
\end{figure}

As already mentioned in Ref.~\cite{Gong2022}, the existence of EPs is not so crucial --- we can still observe the above algebraic decays for $0<J<\kappa/4$. See Fig.~\ref{Fig:NoEP} for an example. Moreover, the detuning to the EP  $\Delta= -i\kappa/2$ does not make a qualitative difference either (although not shown here). What is important here turns out to be the criticality (gaplessness) of the system and the divergent density of states at the eigenvalue with zero imaginary part. Even though for Hermitian systems this algebraic decay is also present, its contribution to the dynamics is obscured by the bound states, which have real energy and therefore do not decay.

An intriguing feature of the alternating-dissipation model described by Eq.~\eqref{eq:alternating-diss} is the fact that the self-energy for emitters coupled to the $A$ sublattice does not diverge, but it vanishes at the branch point $z_\mathrm{BP} = 0$, i.e., $\Sigma_A(0) = 0$. This not only gives rise to the anomalous $t^{-1}$ decay mentioned already, but also to a non-decay eigenstate that may not necessarily have a localized photon profile (since the energy is not well separated from the photon spectrum; cf. Appendix~\ref{App:loc}). Indeed, under PBC, one can check this eigenstate is nothing but the photon Bloch wave created by $\hat a^\dag_{kB}$ with $k=\pi$.

\begin{figure}
    \centering
    \includegraphics[width=0.9\linewidth]{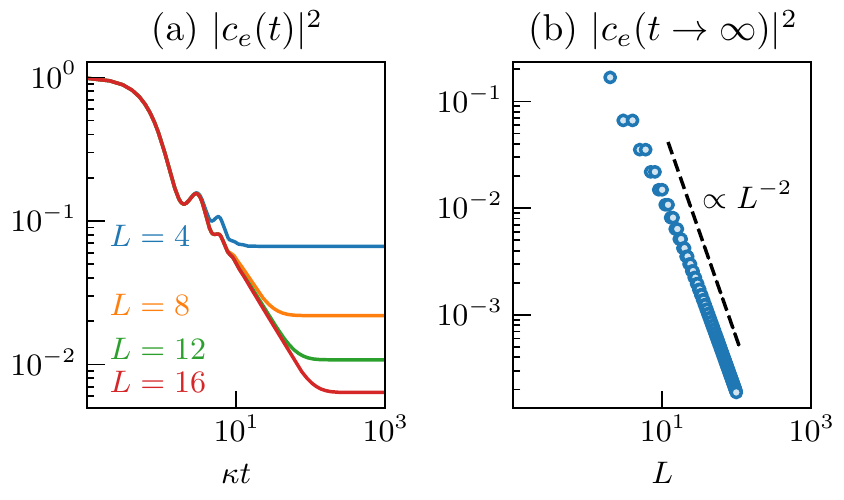}
    \caption{(a) Excited-state population dynamics for a single emitter coupled to an $A$ site in a finite bath of varying size $N$ and OBCs; the emitter is coupled to the middle of the chain. The parameters of the model are $\Delta = 0$, $J=\kappa$, and $g=1.2\kappa$. (b) Long-term value of the excited-state population for the same system.}
    \label{fig:quasiboundstate}
\end{figure}

On the other hand, under OBC and provided that the emitter is close to the right edge as well as $\Delta=0$, there appears a (quasi) bound state in the continuum (BIC). Its wave-function can be chosen to be real, and it is given, up to a normalization constant, by
\begin{equation}
    \ket{\psi_\mathrm{BIC}} \propto \left[\frac{J}{g} \hat{\sigma}^{eg} - \sum_{x \geq x_e} (-1)^{x-x_e} \hat{a}^\dag_{xB}\right]\ket{g}\otimes\ket{\mathrm{vac}} \,, \label{eq:BIC}
\end{equation}
where $x_e$ is the index of the unit cell where the emitter is coupled.
Note that for this model, since $H^{\rm T}_{\rm eff} = H_{\rm eff}$ (under the natural basis consisting of the single atom or photon excitations created by $\hat\sigma^{eg}$ or $\hat a^\dag_{xA/B}$ from $|g\rangle\otimes|{\rm vac}\rangle$), left and right eigenvectors are the complex conjugate of one another.
The consequences of this can be clearly appreciated in Fig.~\ref{fig:quasiboundstate}, where the fractional decay of the initially excited emitter is related to the emitter weight of the quasi-bound state, $\lim_{t\to\infty}\abs{c_e(t)}^2 = \abs{c^e_{\mathrm{BIC}}}^4$, which decreases as $\abs{c^e_{\mathrm{BIC}}}^2 \propto L^{-1}$ as the system size $L$ increases; or rather, as the distance between the emitter and the right end of the chain increases.

BICs of this kind are not unique to NH systems \cite{Tufarelli2013}. Actually, the same state, $\ket{\psi_\mathrm{BIC}}$ in Eq.~\eqref{eq:BIC}, is an eigenstate in the limit $\kappa\to 0$, in which the bath becomes a simple (Hermitian) 1D chain with NN couplings. However, 
there is also curcial difference. Note that in 1D Hermitian systems, BICs similar to the one in Eq.~\eqref{eq:BIC} exist (have non-zero amplitude) between the emitter and \emph{any} edge of the bath, whereas in our NH model stable BICs only exist between the emitter and the right edge of the chain. To prove this, we just have to realize that these BICs are a particular type of vacancy-like bound states~\cite{Leonforte2021}, so their energies coincide with the energies of the eigenstates of open chains. For our model, there are only three different kinds of open chains, depending on the starting and ending sublattice: $AB\dots AB$, $AB\dots BA$, and $BA\dots AB$. It is only this last kind that has an eigenstate with real energy. Of course, a stable BIC with non-zero amplitude between the emitter and the left edge will exist if we allow chains to start with a $B$ site.

\subsubsection{Two-emitter dynamics}

\begin{figure}
    \centering
    \includegraphics[width=0.9\linewidth]{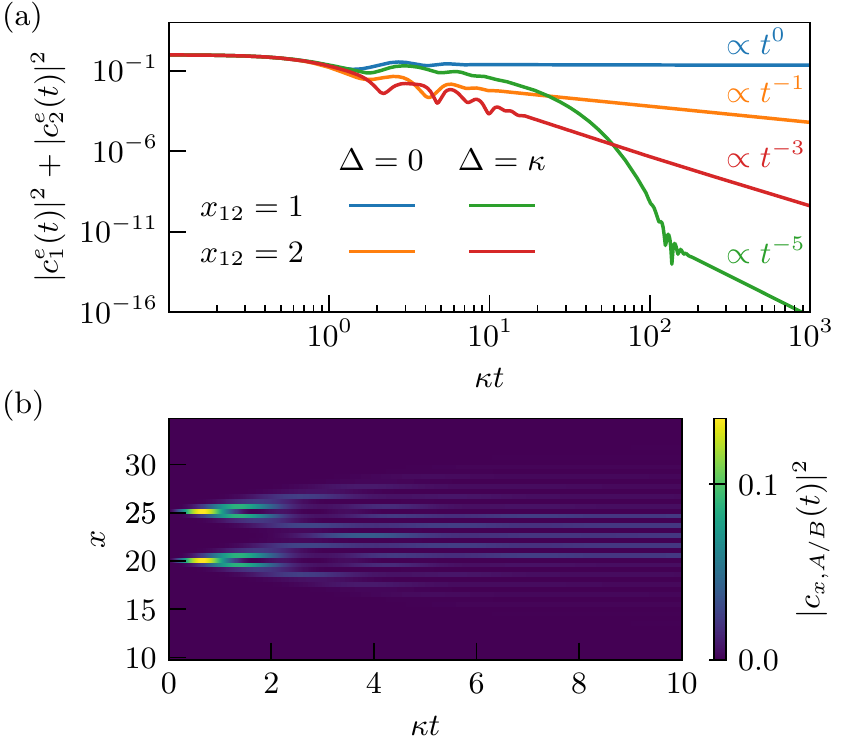}
    \caption{(a) Emitter dynamics for a pair of quantum emitters coupled to the $A$ sublattice. The initial state is the symmetric superposition. The rest of parameters are $J=\kappa$ and $g=1.5\kappa$. (b) Bath dynamics for a system with $x_{12} = 5$ and $\Delta = 0$ [the rest of parameters are the same as in pannel (a)].}
    \label{fig:two_emitter_algebraic}
\end{figure}

For two emitters coupled to the same sublattice, the self-energy is a symmetric two-by-two matrix with constant diagonal elements. Thus, its eigenvectors are the symmetric and antisymmetric superpositions $\boldsymbol{c}_\pm=[1,\pm1]^{\rm T}/\sqrt{2}$ (corresponding to $\ket{\pm} \propto \left(\hat{\sigma}^{eg}_1 \pm \hat{\sigma}^{eg}_2\right)\ket{gg}$), allowing the spectral decomposition:
\begin{equation}
    \Sigma(z) = \Sigma_+(z)P_+ + \Sigma_-(z)P_-\,,
\end{equation}
with $\Sigma_\pm(z) \equiv \Sigma_{11}(z) \pm \Sigma_{12}(z)$ and 
$P_\pm= \boldsymbol{c}_\pm \boldsymbol{c}_\pm^\dag$.  
In other words, the probability amplitudes of the (anti)symmetric superpositions are independent of each other and can be computed the same way as the probability amplitude of the excited state in the single-emitter case, the only difference being the functional form of the self-energy associated to each state. As it turns out, when both emitters are coupled to the $A$ sublattice (dissipative sublattice), depending on the parity of the number of unit cells separating the two emitters, the (anti)symmetric superposition has a nonzero overlap with a proper (normalizable) stable bound state, $\Sigma_-(0)=0$ [$\Sigma_+(0)=0$] if $\abs{x_{12}}$ is even [odd], whose wavefunction has the form (assuming $x_1 < x_2$)
\begin{multline}
    \ket{\psi_\mathrm{b}} \propto \Bigg\{\frac{J}{g}\left[\hat{\sigma}^{eg}_1 + (-1)^{x_{21} + 1} \hat{\sigma}^{eg}_2\right] \\ - \sum_{x_1 \leq x < x_2} (-1)^{x-x_1} \hat{a}^\dag_{xB}\Bigg\}\ket{gg}\otimes\ket{\mathrm{vac}} \,.
    \label{2EB}
\end{multline}
As a consequence, an initial (anti)symmetric state does not fully decay, as shown in Fig~\ref{fig:two_emitter_algebraic}(a) (blue curve), and part of the emitted photon remains trapped indefinitely between the two emitters, see Fig~\ref{fig:two_emitter_algebraic}(b). This proper bound state does not exist when both emitters are coupled to the $B$ sublattice, or to different sublattices. Note that such kind of two-emitter bound state (\ref{2EB}) appears also in the absence of dissipation ($\kappa=0$), as mentioned in Ref.~\cite{Alejandro2017b}.

In addition, with two emitters we can observe other algebraic decays, which are not present in the single-emitter case. For example, for two emitters coupled to the $A$ sublattice out of resonance with the bound state ($\Delta\neq 0$), we find a decay $\propto t^{-5}$, see Fig.~\ref{fig:two_emitter_algebraic}(a) (green curve). Such a power law can again be quantitatively understood from the analytic expression of the self-energy (\ref{PTSE}) and the general formula in Eqs.~(\ref{cBC}) and (\ref{Fnu}) (see Appendix~\ref{App:PTBC} for detail).

\subsection{Two dimensions}
\begin{figure}
	\includegraphics[width=8.5cm, clip]{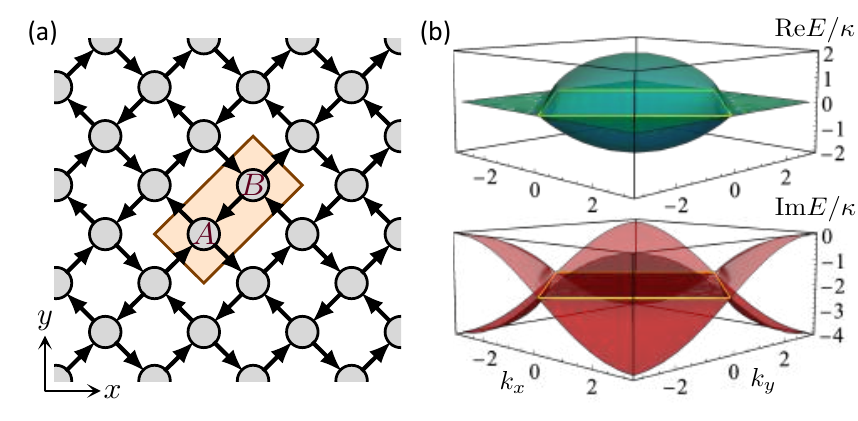}
	\caption{(a) 2D NH lattice described by Eq.~(\ref{2Daf}) (apart from a global loss). Here each arrow represents a unidirectional hopping, and the orange rectangle indicates a unit cell containing two sublattice degrees of freedom $A$ and $B$. (b) Real (upper) and imaginary (lower) band dispersions. The yellow lines indicate the exceptional ring.}
	\label{Fig:2D}
\end{figure}

Finally, let us consider a 2D model with an exceptional ring. As shown in Fig.~\ref{Fig:2D}(a), this model is inspired by the four-step SWAP model \cite{Lindner2013}, which is arguably the simplest example of 2D anomalous Floquet insulators with chiral edge modes but zero Chern number (actually no dynamics in the bulk). The crucial difference here is that each arrow in Fig.~\ref{Fig:2D}(a) is interpreted as a unidirectional hopping rather than shift. Note that the unidirectional Hatano-Nelson model can be obtained from a 1D shift (Thouless pump \cite{Kitagawa2010}) via such a mapping. Recalling that the underlying open system is lossy, we know that the NH Bloch Hamiltonian reads 
\begin{equation}
 h_{\boldsymbol{k}}=\kappa\begin{bmatrix} -2i & 1+e^{-i(k_y + k_x)} \\  e^{ik_x} + e^{i k_y} & -2i \end{bmatrix}.
\label{2Daf}
\end{equation}
This effective NH Hamiltonian can be obtained by choosing $\hat H_{\rm p}$ to involve only NN hopping with amplitude $J=\kappa/2$, $m=4$ and
\begin{equation}
\begin{split}
\hat L^1_{\boldsymbol{r}}= \hat c_{\boldsymbol{r}A} - i\hat c_{\boldsymbol{r}-\boldsymbol{e}_x,B},\;\;\;\;
\hat L^2_{\boldsymbol{r}}= \hat c_{\boldsymbol{r}A} - i\hat c_{\boldsymbol{r}-\boldsymbol{e}_y,B},\\
\hat L^3_{\boldsymbol{r}}= \hat c_{\boldsymbol{r}B} - i\hat c_{\boldsymbol{r}A},\;\;\;\;
\hat L^4_{\boldsymbol{r}}= \hat c_{\boldsymbol{r}B} - i\hat c_{\boldsymbol{r}+\boldsymbol{e}_x+\boldsymbol{e}_y,A}.
\end{split}
\label{L1234}
\end{equation}
Here the sublattice labels $A$ and $B$ are indicated in Fig.~\ref{Fig:2D}(a) and $\boldsymbol{e}_{x,y}$ denotes the unit vector along $x,y$ direction.

\begin{figure}
    \centering
    \includegraphics{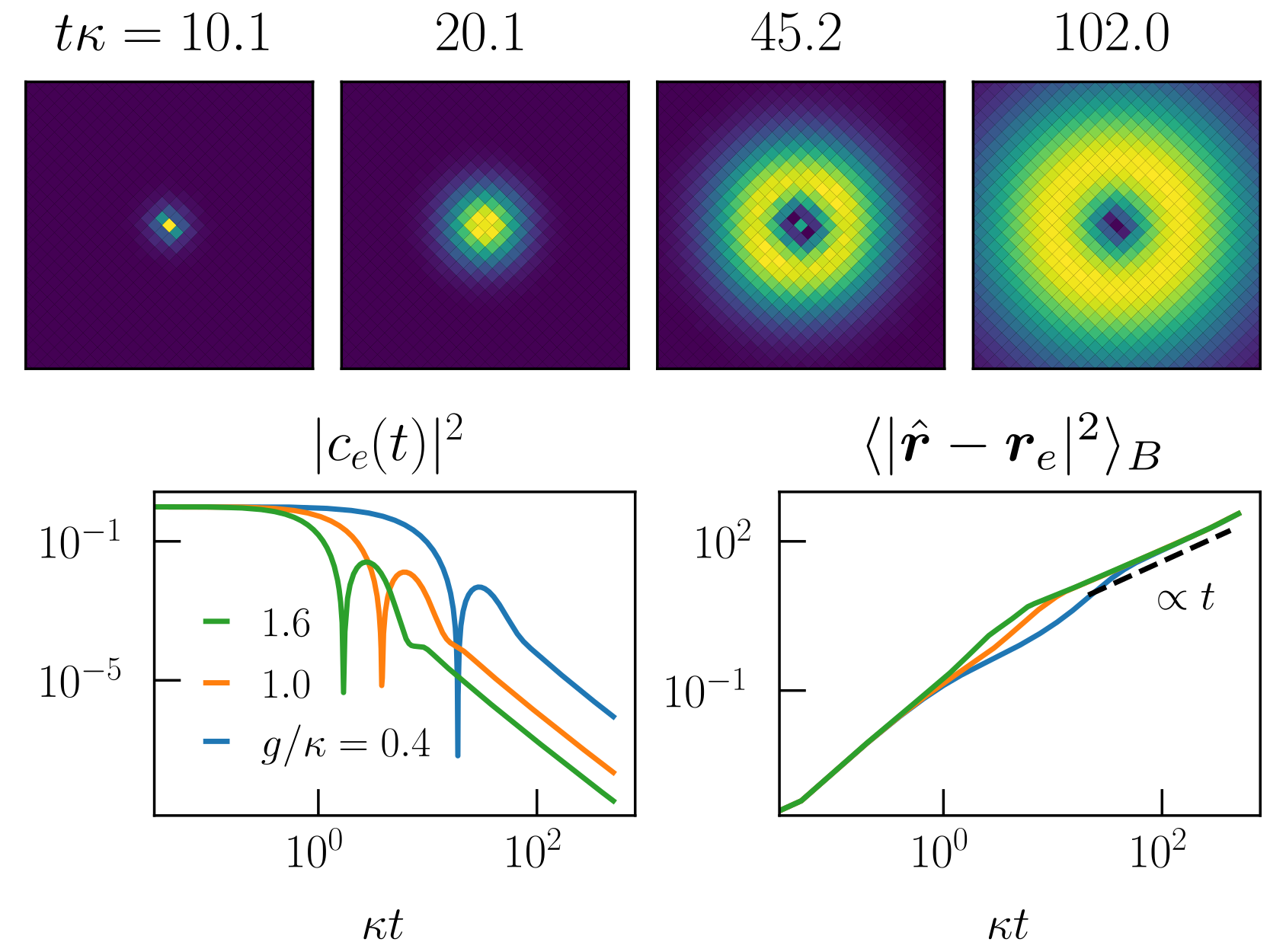}
    \caption{Emission dynamics for a single emitter with $\Delta = 0$, coupled to a single site of the 2D NH lattice described by Eq.~\eqref{2Daf}; the lattice considered has $30\times 30$ unit cells and PBC. (Top row) Bath occupations around the emitter location, $\abs{c_{\boldsymbol{r}}(t)}^2$, at different times, for the case with $g = 0.4\kappa$; the color scale has been normalized in each snapshot such that the brightest color corresponds to the maximum single-site bath occupation. (Bottom row) Emitter excited-state population (left), and mean squared displacement of the photonic cloud (right) as a function of time, for different values of the coupling constant. The expectation value is computed as follows $\mean{\hat{O}}_B\equiv \tr \left[\hat O \hat\rho_B(t)\right]/\tr\left[ \hat\rho_B(t)\right]$, and $\hat{\rho}_B = \hat{P}_B\hat{\rho}\hat{P}_B$, with $\hat P_B = 1 -\hat{\sigma}_0\otimes\ket{\mathrm{vac}}\!\bra{\mathrm{vac}}.$}
    \label{fig:diffusive_bath}
\end{figure}

Interestingly, apart from the background loss, the Bloch Hamiltonian (\ref{2Daf}) is the square root of the NN hopping model:
\begin{equation}
(h_{\boldsymbol{k}} + 2i\kappa\sigma_0)^2=2\kappa^2(\cos k_x + \cos k_y)\sigma_0,
\label{hk2}
\end{equation}
where $\sigma_0$ is the $2\times 2$ identity. Accordingly, one can readily obtain the band dispersions to be 
\begin{equation}
    \epsilon_{\boldsymbol{k}}=-2i\kappa\pm\kappa\sqrt{2(\cos k_x + \cos k_y)},
\end{equation}
implying that the EPs constitute the contour $k_x\pm k_y=\pi\mod 2\pi$ (see Fig.~\ref{Fig:2D}(b)). Using the same property (\ref{hk2}), one can analytically evaluate the single-emitter self-energy to be \cite{Alejandro2017b}
\begin{equation}
\begin{split}
\Sigma(z)&=\int_{\rm B.Z.} \frac{d^2\boldsymbol{k}}{(2\pi)^2} \frac{g^2(z+2i\kappa)}{(z+2i\kappa)^2-2\kappa^2(\cos k_x + \cos k_y)} \\
&=\frac{2g^2}{\pi(z+2i\kappa)}{\rm K}\left[\left(\frac{2\kappa}{z+2i\kappa}\right)^4\right],
\end{split}
\end{equation}
where ${\rm K}(m)\equiv \int^{\pi/2}_0 d\theta/\sqrt{1-m\sin^2\theta}$ is the complete elliptical integral of the first kind. One can check that there is a logarithmic branch point at the origin, around which the self-energy is dominated by \cite{Morita1971}
\begin{equation}
\Sigma(z )\simeq \frac{i g^2}{2\pi \kappa} \ln\left( \frac{z}{8i\kappa}\right).
\end{equation}
Such a singularity is similar to that found in the NN 2D Hermitian model at the band center \cite{Alejandro2017a,Alejandro2017b} and gives rise to a non-exponential atom decay, as numerically confirmed in Fig.~\ref{fig:diffusive_bath}. However, recalling that we are considering the square root of the NN 2D Hermitian model, the observed non-exponential decay should be attributed to the lower branch cut in the Hermitian model, where its contribution is overwhelmed by the lower bound states. In contrast, here this branch point has the largest imaginary part and thus dominates the long-time dynamics.

Quantitatively, for the time interval in our numerical calculations, we find that the non-exponential decay seems to be well described by $\propto t^{-2.5}$, as is also the case in the Hermitian model \cite{Alejandro2017a,Alejandro2017b}. However, by na\"ively identifying the logarithmic scaling on the rhs of Eq.~(\ref{Fnu}) as a zero power ($\nu=0$), we may expect from Eq.~(\ref{cBC}) that the decay follows an algebraic law $t^{-2}$, possibly with logarithmic corrections that may account for the apparent inconsistency. Note that such an ``actual" scaling has also been mentioned in Ref.~\cite{Alejandro2017b}.

In addition, just like the 1D (passive) $PT$-symmetric model (\ref{1DPT}) considered above, the gradient of Eq.~(\ref{2Daf}) at the wave number with the largest imaginary part vanishes. While, unlike the 1D case, this does not lead to a divergent density of states, it significantly alters the photon dynamics, which turns out to be diffusive rather than ballistic, as numerically demonstrated in Fig.~\ref{fig:diffusive_bath}. This result may be understood from the fact that the free photon propagation in this 2D lattice is diffusive. Qualitatively, this may be understood from the fact that at each site the photon can only move along one of the two directions, so the dynamics should be rather constrained. Quantitatively, by expanding the NH Bloch Hamiltonian near the zero damping point $(k_x,k_y)=(\pi,\pi)$, we can estimate the diffusive photon (density) profile to be roughly $(\kappa t)^{-2} e^{-2\boldsymbol{r}^2/(\kappa t)}$ up to a constant coefficient, which also implies the entire photon decay is critical and follows an algebraic law $t^{-1}$.

In fact, the diffusive photon propagation appears also in the passive $PT$-symmetric 1D model (\ref{1DPT}). This property in turn provides an intuition into the algebraic emitter decay observed in all these models: Since the emitted photon propagates very slowly, there is a larger (compared to the ballistic case) probability that it is absorbed (and then emitted) by the emitter again, leading to an enhancement of non-Markovianity. This argument also gives the intuition why the branch-cut contribution becomes large when the detuning is close to the band edges in Hermitian systems \cite{Alejandro2017b}, near which the photon modes have almost vanishing group velocities.

Before ending this subsection, we would like to mention that neither the non-exponential decay nor the diffusive photon dynamics necessarily requires EPs. Indeed, we expect that similar phenomena may be observed in a Wick-rotated 2D Hermitian model (cf. Sec.~\ref{Sec:CSII}). In order to observe exotic phenomena directly related to EPs, we may have to require the imaginary part of EPs to be the largest, a seemingly very uncommon situation. This could be yet another specific open problem for future studies.

\section{Discussions}
\label{Sec:Dis}
In this section, we discuss some general observations we found in our case studies. We also discuss the experimental relevance of our models.

\subsection{Spectral stability}
\begin{figure}
    \centering
    \includegraphics[width=8.5cm, clip]{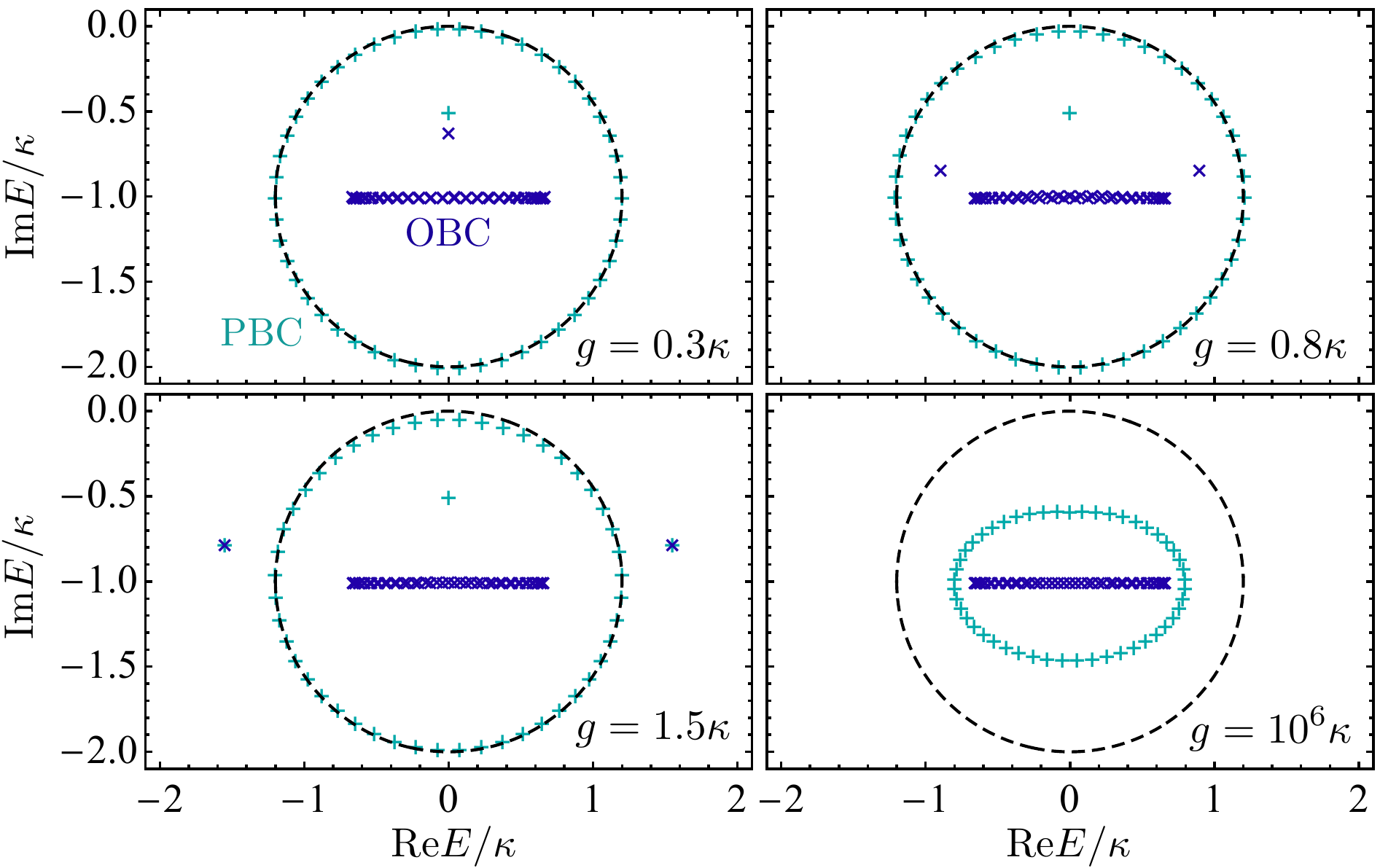}
    \caption{Spectrum of a quantum emitter coupled to a Hatano-Nelson lattice (\ref{HNhk}) with $50$ sites under periodic (green ``$+$") and open (blue ``$\times$") boundary conditions. The bare photon dispersion is indicated by the black dashed curve. In all the panels, $J=0.6\kappa$ and $\Delta=-0.5i\kappa$.}
    \label{fig:BC}
\end{figure}

A unique feature of NH matrices is the spectral instability against perturbations. However, for all the examples considered above, where we put one or a few emitters in periodic lattices in the thermodynamic limit, the system spectra do not seem to be altered so much except for the bound states (see Fig.~\ref{fig:BC} for an explicit illustration). It is thus natural to ask whether the spectral stability is a universal property. For simplicity, we focus on the case of a single emitter.

 One can 
show that the number of bound states outside 
the bare photon spectrum (especially excluding BICs) should always be finite in the thermodynamic limit. We first recall that the eigenenergies of bound states are determined by Eq.~(\ref{det0}), i.e., the poles of the Green's function. Provided that the system is local, we know that the Bloch Hamiltonian $h_{\boldsymbol{k}}$ depends smoothly on $\boldsymbol{k}$ and so will the self-energy $\Sigma(E)$ on $E$ away from the bare photon spectrum. In complex analysis, this property is known as analyticity, implying that the function is ``holographic", i.e., it can be fully determined from a small region, or even a subset of infinite points within it. In particular, if an analytic function has infinite zeros on a finite region, it is identically zero \cite{Ahlfors1979}. Now back to the lhs of Eq.~(\ref{det0}), which is analytic in $E$ and its zeros should locate in a bounded region since $|\Sigma(E)|$ decay like $|E|^{-1}$ for large $|E|$, we know that an infinite number of bound states implies that the lhs of Eq.~(\ref{det0}) is identically zero, leading to a contradiction. 


The remaining problem is whether dramatic spectral change could occur on and inside 
the bare photon spectrum, such as a significant redistribution of the density of states. 
We conjecture this cannot occur either, provided that we keep $g$ finite or take the thermodynamic limit first. The intuition is that, according to the finite-size version of Eq.~(\ref{det0}) which applies to any eigenstates, there should be a solution near each eigenvalue of $h_{\boldsymbol{k}}$, around which the self-energy becomes very sensitive. To explain a bit more why this problem could be difficult, let us focus on 
a single-band lattice with band dispersion $h_{\boldsymbol{k}}$. If the system is Hermitian and the detuning is real, one can readily check that Eq.~(\ref{det0}) is equivalent to $V'(E)=0$ with
\begin{equation}
V(z)=\frac{|\Lambda|}{2g^2}(z-\Delta)^2-\sum_{\boldsymbol{k}\in{\rm B.Z.}}\ln|z-h_{\boldsymbol{k}}|.
\label{Vz}
\end{equation}
This function may be interpreted as a potential for a charged particle arising from a harmonic trap centered at $\Delta$ and an array of charged particles fixed at $h_{\boldsymbol{k}}$'s. Here the Coulomb force is repulsive and proportional to the inverse of distance. The zeros of $V'(z)$ are thus (stable) equilibrium points of this potential. Intuitively, there should be equilibrium points between any two adjacent charges. 
This result suggests a small deviation from the bare photon spectrum. For NH systems, however, we should get rid of the absolute values in the Coulomb potentials in Eq.~(\ref{Vz}). Hence, $V(z)$ becomes complex in general and can no longer be interpreted as a potential. It is thus not obvious whether the zeros of $V'(z)$ are guaranteed to be around $h_{\boldsymbol{k}}$'s. Mathematically, the crucial difference is related to the fact that 
the mean value theorem, which holds true for real functions, generally breaks down for complex functions \cite{Qazi2006}. 

On the other hand, if we first take the large-$g$ limit for a finite-size system, 
we expect the system spectrum would be that of the NH lattice with some vacancy defects at the sites directly coupled to the emitters \cite{Leonforte2021}. This may significantly alter the spectrum (see the right-bottom panel in Fig.~\ref{fig:BC} for an example) due to the boundary condition sensitivity of NH systems, as will be discussed in the next subsection. Note that the non-commutativity between thermodynamic limit and certain limit of coupling strength has already been highlighted in Refs.~\cite{Gong2018,Okuma2019} for NH lattices alone.

\subsection{Impact of boundary conditions}
As an important implication of the instability of NH matrices against perturbations, it is known that both spectra and eigenstates of NH systems may dramatically change under different boundary conditions. Indeed, such boundary condition sensitivity appears in any NH system with skin effect, exemplified by the Hatano-Nelson model. This is also the case in the presence of emitters. For example, for the Hatano-Nelson model (\ref{HNhk}), one can check that while all bound states are localized near the emitter under PBC, all the eigenstates are localized at one boundary under the OBC, provided that the coupling strength is not so strong. This applies particularly to those bound-state like modes (whose eigenvalues obviously locate outside the ``bulk" spectrum under the OBC) in the upper panels in Fig.~\ref{fig:BC}.

On the other hand, we found that the atom-decay and photon-emission dynamics do not depend on the boundary condition, at least in the short-time regime. In fact, it has recently been proved for an arbitrary NH lattice with finite-range hopping that the wave propagation in the bulk is (almost) not altered by the boundary condition \cite{Mao2021}. The proof applies directly to our setups and justifies the boundary insensitivity of bulk dynamics in general. Here, we provide an alternative argument based on the equivalence of Lindblad dynamics and NH Hamiltonian evolution in the single-particle sector of a lossy Markovian open system. By further imposing locality, we know that the Lindblad dynamics satisfies the Lieb-Robinson bound \cite{Poulin2010}, and thus so does the NH Hamiltonian evolution. This further implies that the dynamics in the bulk should have weak boundary-condition dependence \cite{Wang2021}. 

It is worthwhile to mention that the Lieb-Robinson bound is expected to be generally violated in NH systems with multiple particles \cite{Ashida2018,Matsumoto2020}. An intuitive understanding is that non-local measurements and post-selections are usually required to achieve NH Hamiltonian evolutions. This observation may potentially imply nontrivial boundary dependence for dynamics in NH nanophotonic systems upon going beyond the single-particle paradigm.


\subsection{Validity of the single-pole approximation}
Perhaps the simplest approach to understanding the emitter dynamics in Hermitian baths is the single-pole approximation (SPA), which consists of replacing the self-energy $\Sigma(E)$, appearing in the integral \eqref{eq:cet}, by a constant $\Sigma(\Delta)$. Such an approximation is also known as the Fermi's golden rule or the Wigner-Weisskopf approach \cite{Sakurai2011}. In this approximation, the effect of the whole bath is reduced to a shift (possibly with both a real and an imaginary component) of the emitter detuning: 
\begin{equation}
c_e(t)\simeq e^{-i[\Delta +\Sigma(\Delta)]t}.
\label{SPA}
\end{equation}
This is valid in the weak-coupling regime ($g\ll\kappa,J$) and when the self-energy $\Sigma(E)$ is smooth around $E=\Delta$. 

Within this approximation it is easy to understand that for Hermitian systems there is a transition in the emitter dynamics: the excited state population either decays or not depending on whether the emitter is resonant or not with the bath modes. At the transition between the two kinds of behavior the self-energy is singular, so the SPA breaks down---it is where the non-Markovian effects are the strongest.

\begin{figure}
    \centering
    \includegraphics[width=8.5cm, clip]{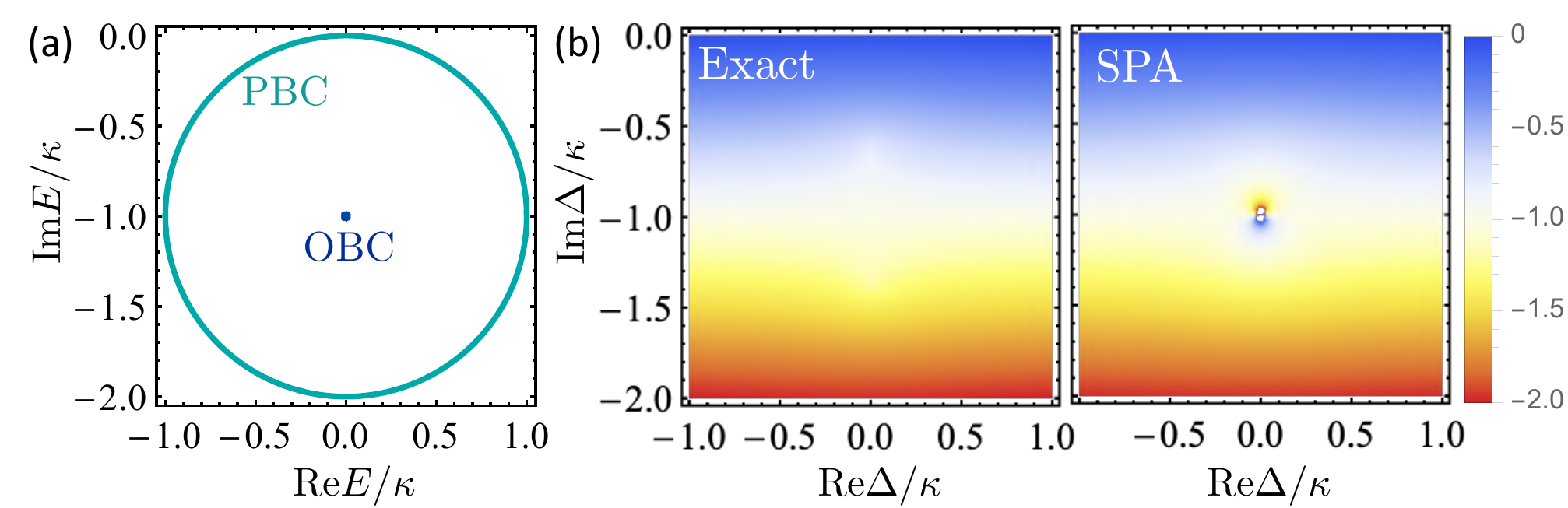}
    \caption{(a) Spectrum of the unidirectional Hatano-Nelson model under PBC (green) and OBC (blue). (b) Dependence of the imaginary part of the pole (divided by $\kappa$) on the complex detuning $\Delta$ for $g=0.2\kappa$. The left and right panels show the exact results (\ref{exact}) and those under the SPA (\ref{SPA}), respectively. Clearly, significant difference appears only near the OBC spectrum.}
    \label{fig:SPA}
\end{figure}

One may wonder whether a similar effect occurs in NH systems. Is there a dramatic change in the emitter dynamics when the emitter is resonant with the bath modes? The answer is not so straightforward as in the Hermitian case, since the spectrum for OBC or PBC can be very different (especially in NH systems with point-gap topology). As it turns out, being resonant with PBC modes may not have a big impact on the emitter dynamics. For example, in Fig.~\ref{fig:SPA} we plot the decay rate for a single emitter coupled to the unidirectional Hatano-Nelson model, i.e., Eq.~(\ref{HNhk}) with $J=\kappa/2$. The emitter dynamics is given exactly by 
\begin{equation}
c_e(t) = R_+ e^{-iz_+t} + R_- e^{-iz_-t},
\label{exact}
\end{equation}
where $2 z_\pm = \Delta - i\kappa \pm \sqrt{(\Delta + i\kappa)^2 + 4g^2}$ and $R_\pm = \pm (z_\pm + i\kappa)/(z_+ - z_-)$. The ``exact'' decay rate, which is taken as the imaginary part of the pole that is closest to $\Delta$ ($z_+$ if $\Re \Delta > 0$, or $z_-$ otherwise), agrees well with the one given by the SPA all over the PBC spectrum. By contrast, they disagree when $\Delta$ is tuned close to the OBC spectrum, where, in fact, the self-energy diverges. Actually, for $\Delta$ tuned to the singularity ($\Delta = -i\kappa$) $\abs{z_+ - \Delta} = \abs{z_- - \Delta}$ and $\abs{R_+}=\abs{R_-}$, so in that case the SPA breaks down, no matter how small $g$ is.

Let us provide a plausible argument about why the breakdown of SPA is more relevant to 
the spectrum under OBC instead of PBC in 1D. In general, a 1D NH Bloch Hamiltonian $h_k$ may exhibit some nontrivial spectral winding topology so that its spectrum forms one or several loops. The self-energy can be evaluated as (cf. Eq.~(\ref{Sz}))
\begin{equation}
  \Sigma(z)=\oint_{|\beta|=1} \frac{d\beta}{2\pi i \beta} g^\dag_\beta (z-h_\beta)^{-1}g_\beta, 
\end{equation}
where we have replaced $k$ by $\beta\equiv e^{-ik}$. Recalling that the NH lattice is assumed to be short-ranged and the atom-photon coupling is on-site, we know that $h_\beta$ and $g_\beta$ are analytic (on the whole complex plane excluding $\beta=0,\infty$) and so is $\Sigma(z)$ outside the loop. As long as the spectrum of $h_\beta$ does not touch $z$, one can freely deform the integral contour to a circle with arbitrary radius, which actually corresponds to the twisted boundary condition with an imaginary flux \cite{Hatano1996,Hatano1997}, while keeping the integral result invariant. This also gives a natural way of ``holographically" reconstructing 
$\Sigma(z)$, technically known as analytic continuation, since the spectrum of $h_\beta$ might shrink on a deformed contour. Denoting $D$ and $D'$ as the interiors of the original and shrinked spectra, we know that $\Sigma(z)$ can be analytically extended to $D\setminus D'$ via the deformation. In fact, it has been shown that the common part (set intersection) of the interior of the spectra of $h_\beta$ for various contours gives nothing but the OBC spectrum \cite{Okuma2020}. This implies that the analytic continuation of $\Sigma(z)$ is possible outside the OBC spectrum. Also, the analytically continued $\Sigma(z)$ necessarily exhibits some singularities (divergence) on the OBC spectrum, where perturbative analysis in terms of small $g$ breaks down and so does the SPA.

\subsection{Experimental implementations}
Many experimental platforms exist in which non-Hermititian lattice Hamiltonians (see for example the overviews in Refs.~\cite{Kunst2021,Ashida2021}), or even the full Lindbladian dynamics (see, e.g., Refs.~\cite{Poyatos1996,Kraus2008,Muller2012}) can be realized.
In the systems considered in this paper, one or multiple additional modes or qubits (emitters) have to be coupled to the lattice.
To make contact with the typical set-up in quantum optics, the emitters need to effectively be two-level systems, such as atoms or qubits. While this is not required to study the single-particle effects that we discuss here, it will be crucial to observe many-particle effects such as Dicke superradiance.
Closest to the idea of emitters coupled to NH baths are realizations such as atoms coupled to nanophotonic structures~\cite{Chang2018}. In such devices, loss of photons in the bath naturally occurs, but it may be difficult to control. Better control is afforded in synthetic platforms, such as superconducting qubits coupled to superconducting metamaterials~\cite{You2011,Kim2021}, or ultracold atoms in optical lattices (see proposal in Ref.~\cite{Tao2016}).
Since all of the physics discussed here is single-photon physics, having two-level emitters is not strictly necessary, and any synthetic platform in which NH arrays can be (weakly) coupled to isolated modes constitutes a viable platform.
In the following, we describe a concrete idea how collective loss may be engineered in synthetic platforms.

\begin{figure}
	\includegraphics[width=8.5cm, clip]{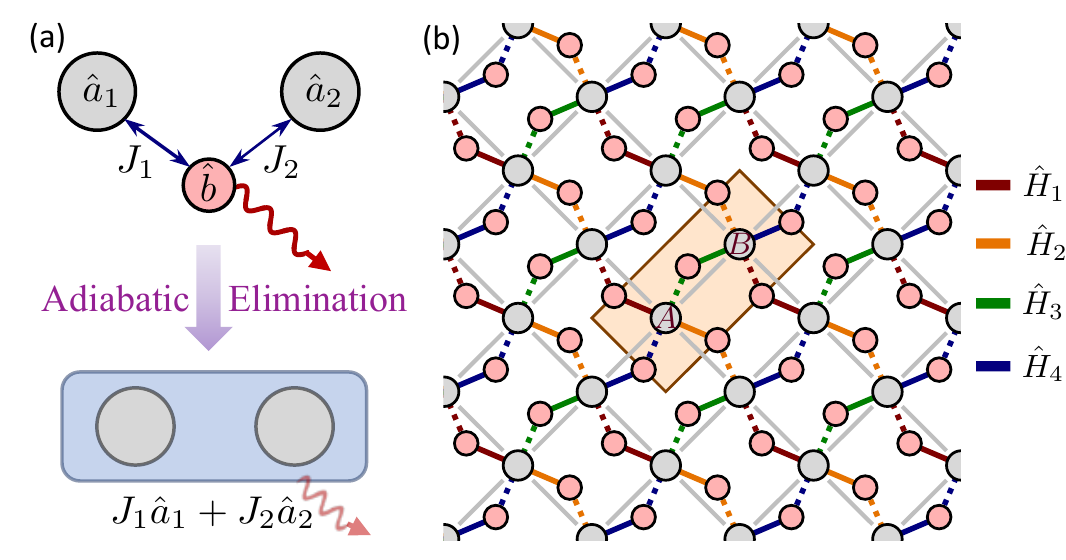}
	\caption{(a) Schematic illustration of realizing collective loss of two modes $\hat a_1$ and $\hat a_2$ by coupling them to a common ancilla mode $\hat b$. If the ancilla mode has a large decay rate, we can adiabatically eliminate it, obtaining an effective collective loss for  $\hat a_1$ and $\hat a_2$.  (b) Implementation of the 2D NH model in Eq.~(\ref{2Daf}). Here the gray (pink) circles correspond to the primary (auxiliary) degrees of freedom. $\hat H_\mu$ ($\mu=1,2,3,4$) is responsible for the jump operator $\hat L^\mu$ in Eq.~(\ref{L1234}). Solid and dotted bonds in color share the same coupling strength but differ by a phase factor $i$.}
	\label{Fig:Exp}
\end{figure}

A key challenge in any platform is to introduce spatially non-local collective loss as described by Eq.~(\ref{Lmr}). A general idea is to introduce some auxiliary rapidly decaying modes and couple them coherently and non-locally to the primary degrees of freedom. After adiabatically eliminating these fast modes \cite{Sorensen2012}, we will end up with a collective loss of the primary degrees of freedom. This technique has been described in detail in Ref.~\cite{Metelmann2015,Ranzani2015,Sounas2017} and demonstrated experimentally in coupled resonators~\cite{Estep2014}, Josephson junctions~\cite{Sliwa2015} and optomechanics~\cite{Fang2017,Bernier2017}. See Fig.~\ref{Fig:Exp}(a) for an illustration for the minimal two-mode setup. In general, if we can introduce a set of auxiliary degrees of freedom $b_{\boldsymbol{r}\mu}$ on the same lattice, where $\mu=1,2,...,m$ and each internal state $\mu$ undergoes on-site loss with decay rate $\kappa_n$, we can engineer the desired collective loss operators $\hat L^\mu_{\boldsymbol{r}}$'s by coupling these auxiliary modes to the primary photon modes of interest through 
\begin{equation}
\hat H_\mu = \sum_{\boldsymbol{r},\boldsymbol{r}'\in\Lambda}\sum_{s'\in I} (\sqrt{\kappa_\mu\kappa} l^\mu_{\boldsymbol{r},\boldsymbol{r}'s'}\hat b^\dag_{\boldsymbol{r}\mu} \hat a_{\boldsymbol{r}'s'} + {\rm H.c.}),
\end{equation}
provided that $\kappa_\mu$ is large enough compared to other parameters such that the adiabatic elimination is well justified. Here, we provide a concrete example of implementing the 2D model in Eq.~(\ref{2Daf}). As shown in Fig.~\ref{Fig:Exp}(b), there are four auxiliary modes (pink circles) associated with one unit cell, each coupled to two nearest primary modes (gray circles). To realize the jump operators in Eq.~(\ref{L1234}), we should fine-tune the couplings between the auxiliary and primary modes indicated by the solid bonds in Fig.~\ref{Fig:Exp}(b) to be $J_{\rm ap}$, while those indicated by the colored bonds to be $-iJ_{\rm ap}$.

Let us provide a bit more detail about how to realize the models we studied on specific experimental platforms. We will specifically focus on platforms that directly extend to the many-body regime, which requires non-linear emitters.
Ultracold atoms in optical lattices have been proposed as a candidate for realizing Hermitian nanophotonic systems~\cite{DeVega2008} and NH lattices~\cite{Ashida2017,Gong2018,Liu2019}.
Following these proposals, a localized emitter decaying into a bath can be simulated by employing two (meta)stable states of a bosonic atom $\ket{s}$ and $\ket{g}$, trapped in a state-dependent optical lattice that is very deep for $\ket{s}$ and shallow for $\ket{g}$. In this configuration, the atom is localized if it is in $\ket s$ (corresponding to an excited emitter) but itinerant if it is in state $\ket g$ (corresponding to a photon). Using a two-photon transition via an excited state driven by two lasers, the stationary atom in $\ket s$ can be coupled coherently to itinerant atoms at a specified frequency~\cite{DeVega2008}.
Single-site addressing can be employed to flexibly engineer the positions of the emitters. The hopping amplitudes with complex phases may be engineered by using techniques developed for generating artificial gauge fields~\cite{Dalibard2011,Goldman2014}. When going to the many-body regime, we can use Feshbach resonances~\cite{Bloch2008} to set the interactions among the photons $\ket g$ and between photons $\ket g$ and emitters $\ket s$ to zero, while enforcing an effective two-level nonlinearity by making the interaction among $\ket s$ large.
Finally, spatially dependent loss can be introduced via another Raman transition to an untrapped state.

Superconducting circuits serve as yet another ideal platform. Here,
the two main ingredients could be transmon qubits~\cite{Koch2007} and a superconducting metamaterial, as demonstrated in Ref.~\cite{Mirhosseini2018}. To control the interactions between modes, tunable couplers can be used, which have been demonstrated in many different settings (see Ref.~\cite{Stehlik2021} and references therein).
Non-reciprocal coupling requires an effective gauge field in addition to dissipation, which can be achieved through parametric driving~\cite{Fang2012a}. Passive implementations have also been realized~\cite{Muller2018}.
This control can then be used to directly realize the scheme in Fig.~\ref{Fig:Exp}; this has been demonstrated for few modes already~\cite{Estep2014,Sliwa2015}.

\begin{figure}
    \centering
    \includegraphics[width=.9\linewidth]{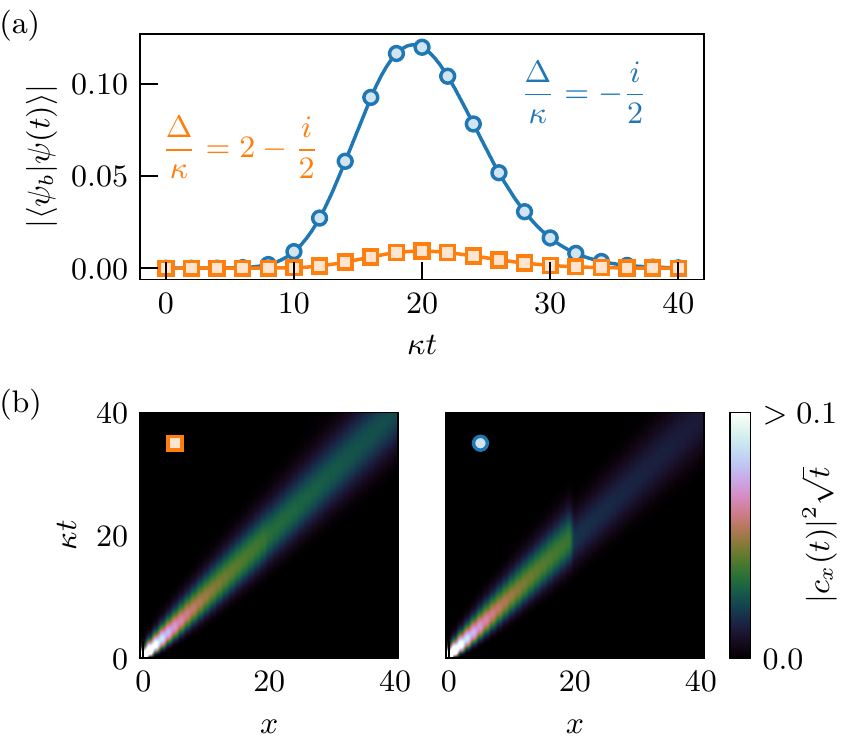}
    \caption{We simulate an emitter coupled with strength $g=0.5\kappa$ to a fully directional Hatano-Nelson chain (Eq.~\eqref{HNhk}, with $J=\kappa/2$) with a total of $L=80$ sites and periodic BCs.
    The initial state consists of a single excitation localized at the 0th bath site, while the emitter is coupled at the 20th bath site.
    (a) Overlap of the time-evolved wavefunction with the bound state at different detunings $\Delta=(2-i/2)\kappa$ (orange squares) and $\Delta=-i\kappa/2$ (blue dots). In the former (latter) case, there is only a Hermitian-like (hidden) bound state.  
    Note that for a Hermitian bath, the overlap would be constant (cf. Eq.~\eqref{eq:overlap}).
    (b) Probability to find the excitation in different bath sites as a function of time.}
    \label{fig:exciting_BIC}
\end{figure}

Before ending this subsection, we would like to discuss the possibility of exciting bound states in NH nanophotonic systems, a topic with particular experimental interest. In the context of waveguide quantum electrodynamics, 
particular attention has been devoted to BICs, 
which may arise when retardation in the waveguide is no longer negligible~\cite{Calajo2016}.
From the point of view of applications, an important question is how this state can be addressed 
and detected. 
Since a BIC is an eigenstate of the Hermitian emitter--waveguide Hamiltonian in the single-excitation subspace (e.g., Eq.~(\ref{eq:BIC}) in the zero dissipation limit), it cannot be excited by a single-photon wavepacket, but instead requires carefully engineered two-photon wavepackets~\cite{Calajo2019}. To reach optimal excitation probability, very long pulses have to be used~\cite{Trivedi2020}.
The reason why single-photon wavepackets do not suffice can easily be understood by calculating the time-dependent overlap of a wavepacket $\ket{\psi(t)}$ with the bound-state wavefunction $\ket{\psi_{\rm b}}$: 
\begin{equation}
    \bra{\psi_{\rm b}}e^{-i \hat H_\mathrm{Hermitian} t}\ket{\psi(0)}=e^{-iE_{\rm b}t}\langle \psi_{\rm b}|\psi(0)\rangle, \label{eq:overlap}
\end{equation}
where $\hat H_\mathrm{Hermitian}|\psi_{\rm b}\rangle=E_{\rm b}|\psi_{\rm b}\rangle$. In contrast, a key feature of NH systems is that left- and right-eigenstates of the Hamiltonian are distinct and that eigenstates are no longer necessarily orthogonal and thus Eq.~\eqref{eq:overlap} ceases to hold.
As a result, the overlap between the bound state $\ket{\psi_{\rm b}}$ and the time-evolved state $\ket{\psi(t)}$ is no longer conserved and therefore, in stark contrast to Hermitian systems, the bound state \emph{can} be excited from the continuum. 
We illustrate this fact with a simple example involving one emitter coupled to a Hatano-Nelson chain, see Fig.~\ref{fig:exciting_BIC}. Interestingly, while both Hermitian-like and hidden bound states can be excited as the simple argument above always applies, it turns out that the latter can be excited more efficiently, at least for our specific choices of parameters. A more qualitative analysis may require a framework for studying scatterings \cite{Zhou2008} between photons travelling in NH baths and emitters. This goes beyond the scope of this paper and we would like to leave it as a future work.

We note that the (complex) overlap $\langle\psi_{\rm b}|\psi(t)\rangle$ is a linear combination of the complex amplitudes of the state on each site. On photonic platforms this can be measured using homodyne tomography~\cite{Lvovsky2009}. In a classical platform such as a mechanical metamaterial~\cite{Ghatak2020} this can directly be obtained from the trajectory of each resonator.


\section{Conclusion and outlook}
\label{Sec:CO}

In summary, we have established a general framework for exploring NH physics in nanophotonic systems with engineered dissipation. We have also provided some case studies on a couple of minimal models that nevertheless exhibit rich phenomena. In particular, we have demonstrated that the hidden bound states (of multiple emitters) and unconventional photon-emission dynamics unveiled in Ref.~\cite{Gong2022} appeared also in the general Hatano-Nelson model. We have pointed out that an algebraic atom decay can be readily achieved by an imaginarily (Wick) rotated 1D lattice, and new exponents appear for multiple emitters in the passive $PT$-symmetric lattice with alternating loss. We have also studied a NH 2D model with non-exponential emitter decay and diffusive photon dynamics. Finally, we have discussed some general features and possible experimental realizations of these NH nanophotonic systems.

There are a lot of possible directions for future studies. Aside from the specific problems mentioned in the main text, the many-body generalization \cite{Tao2016,Asenjo2017,Mahmoodian2020}, where atomic (spin) and photonic (bosonic) excitations are no longer equivalent, could be a natural but challenging project. In this context, a question of particular interest is whether the unique NH features based on the single-particle picture, such as the point-gap topology, still play an essential role. On top of the ultrastrong-coupling regime \cite{Nori2019,Solano2019,Eduardo2019} 
mentioned in Ref.~\cite{Gong2022}, we would like to point out another relevant situation in which the bath involves some parametric amplifications described by pairing terms like $\hat a \hat a + {\rm H.c.}$. Note that such a bath alone may exhibit some genuine NH topology even in the absence of dissipation \cite{Clerk2018}, although its influence on quantum emitters remains unexplored. 

As stated in Ref.~\cite{Gong2022}, even on the single-particle level there are many natural generalizations to, e.g., giant-atom emitters with spatially non-local couplings \cite{Kockum2018,Alejandro2019,Andersson2019}, other NH band topology \cite{Gong2018,Lee2019a,Kawabata2019a,Okuma2020,Kunst2020} and exceptional structures \cite{Hodaei2017,Zhou2019,Mandal2021,Delplace2021,Kunst2022}, and to systems with disorder \cite{Clara2021}, which are ubiquitous in real experiments. One may also consider how the bath-mediated interaction between the emitters, which is already nontrivial for atomic arrays in the vacuum \cite{Asenjo2017,Perczel2017,Perczel2017a,Brechtelsbauer2021}, could be enriched by introducing ``unnatural" non-Hermiticity. 
Further studies along this line may also open up new possibilities for practical applications in, e.g., quantum simulation \cite{Javier2019}, quantum state preparation \cite{Alejandro2015,Seetharam2022}, and quantum metrology \cite{Paulisch2019,Bai2019}.

Finally, we would like to mention two specific problems that are, in our opinions, of particular interest and require systematic studies. One is to classify various van Hove singularities in NH systems and clarify their impact on emitter dynamics. To make a connection with branch cuts in the Green's function, we expect it is more appropriate to consider the singularities associated with the GBZ. In 1D, these singularities should appear at the ends of the tree-like (OBC) energy spectra. The other problem is to study
quantum emitters located at the boundaries of Hermitian topological systems (e.g., Chern insulators). Inspired by the Hermitian-NH correspondence discovered in Ref.~\cite{Lee2019b}, we expect such a setting may share some similarities with that in which emitters are embedded in the bulk of certain NH topological systems. For example, the emitter dynamics at the edge of a 2D Quantum Hall (Chern) insulator should be comparable to that in the bulk of the Hatano-Nelson model, as studied in this paper.

\acknowledgements
We acknowledge Ignacio Cirac and Yuto Ashida for helpful discussions. Z.G.\ is supported by the Max-Planck-Harvard Research Center for Quantum Optics (MPHQ). F.K.K.\ was supported by the Max-Planck-Harvard Research Center for Quantum Optics (MPHQ) before moving to MPL. M.B.\ and D.M.\ acknowledge funding from the ERC Advanced Grant QUENOCOBA under the EU Horizon 2020 program (Grant Agreement No. 742102).

\appendix

\section{Exponential localization of the photon profile}
\label{App:loc}
We have shown in the main text that for a bound state $\boldsymbol{c}_{\boldsymbol{k}}$ can be related to $\boldsymbol{c}_e$ via Eq.~(\ref{ckce}) and then $\boldsymbol{c}_{\boldsymbol{r}}\equiv[c_{\boldsymbol{r}s}]^{\rm T}_{s\in I}$ can be obtained by Eq.~(\ref{crs}). Combining these two equations allows us to express $\boldsymbol{c}_{\boldsymbol{r}}$ explicitly as $\boldsymbol{c}_{\boldsymbol{r}}= \sum^N_{n=1} \boldsymbol{c}_{\boldsymbol{r}n}$, where
\begin{equation}
\boldsymbol{c}_{\boldsymbol{r}n}= c^e_n\int_{\rm B.Z.}\frac{d^d\boldsymbol{k}}{(2\pi)^d}\frac{e^{i\boldsymbol{k}\cdot(\boldsymbol{r}-\boldsymbol{r}_n)}}{E-h_{\boldsymbol{k}}}[g_{ns}]^{\rm T}_{s\in I}.
\end{equation}
Provided that $E$ is outside the spectrum of $h_{\boldsymbol{k}}$ $\forall\boldsymbol{k}\in{\rm B.Z.}$, there should exist $\boldsymbol{K}\in\mathbb{R}_+^d$ such that $(E-h_{\boldsymbol{k}})^{-1}$ is analytic on $D=\{\boldsymbol{k}: \Re\boldsymbol{k}\in{\rm B.Z.},|\Im\boldsymbol{k}|\le \boldsymbol{K} \}$, no matter whether $h_{\boldsymbol{k}}$ is Hermitian or not \cite{Kato1966}. Deforming the integral contour, we can express $\boldsymbol{c}_{\boldsymbol{r}n}$ as
\begin{equation}
\boldsymbol{c}_{\boldsymbol{r}n}=c^e_n\int_{\rm B.Z.}\frac{d^d\boldsymbol{k}}{(2\pi)^d}\frac{e^{i(\boldsymbol{k}+i\boldsymbol{K}')\cdot(\boldsymbol{r}-\boldsymbol{r}_n)}}{E-h_{\boldsymbol{k}+i\boldsymbol{K}'}}[g_{ns}]^{\rm T}_{s\in I},
\end{equation}
as long as $|\boldsymbol{K}'|\le\boldsymbol{K}$. In particular, we can choose $\boldsymbol{K}'$ such that $|\boldsymbol{K}'|=\boldsymbol{K}$ while the sign of each spatial component is the same as that of $\boldsymbol{r}-\boldsymbol{r}_n$. In this case, we can bound the vector norm of $\boldsymbol{c}_{\boldsymbol{r}n}$ from above by
\begin{equation}
\|\boldsymbol{c}_{\boldsymbol{r}n}\|\le C_n e^{-\boldsymbol{K}'\cdot(\boldsymbol{r}-\boldsymbol{r}_n)},
\label{crn}
\end{equation}
where $C_n=|c^e_n|M \sqrt{\sum_{s\in I}g^2_{ns}}$ with $M\equiv\max_{\boldsymbol{k}\in D} \|(E-h_{\boldsymbol{k}})^{-1} \|$ does not depend on $\boldsymbol{r}$. It is now clear from Eq.~(\ref{crn}) that the photon profile should be exponentially localized near the atoms.

We note that the idea of analytically extending $\boldsymbol{k}$ dates back to the demonstration of Wannier localization in 1D Hermitian lattices over sixty years ago \cite{Kohn1959}. Here we emphasize that this idea applies equally to NH systems. Moreover, unlike the Wannier state which may not be exponentially localizable in two and higher dimensions due to, e.g., a nontrivial Chern number \cite{Marzari2007}, the bound state is always exponentially localized, just like the correlation in the ground state of a gapped phase with local interactions \cite{Hastings2006}. For free-fermion systems, this difference can be understood from the fact that, even if the Bloch projector onto the Fermi sea is continuous in $\boldsymbol{k}$ \cite{Cloizeaux1964}, one may not be able to find a set of (filled) Bloch states that are also continuous in $\boldsymbol{k}$ due to a topological obstruction such as a nontrivial Chern number.

\section{Resolvent method}
\label{App:res}
We consider lossy dynamics generated by a NH Hamiltonian $\hat H$ with no eigenvalue above the real axis in the complex energy plane. Suppose that the initial state $|\psi_0\rangle$ is inside a subspace onto which the projector is $\hat P$, i.e., $\hat P|\psi_0\rangle=|\psi_0\rangle$, we can decompose $|\psi_t\rangle=e^{-i\hat Ht}|\psi_0\rangle$ into $\hat P|\psi_t\rangle+\hat Q|\psi_t\rangle$ ($\hat Q\equiv1-\hat P$ is the complement of $\hat P$). The component $\hat P|\psi_t\rangle$ can be evaluated by
\begin{equation}
\begin{split}
&\hat P|\psi_t\rangle=\hat Pe^{-i\hat Ht}\hat P|\psi_0\rangle \\
=&\frac{i}{2\pi}\int^\infty_{-\infty}d\omega \hat G_P(\omega+i0^+)e^{-i\omega t}|\psi_0\rangle,
\end{split}
\end{equation}
where $\hat G_P(z)$ is the constrained propagator defined as
\begin{equation}
\hat G_P(z)\equiv \hat P(z-\hat H)^{-1}\hat P.
\end{equation}
Likewise, the component $\hat Q|\psi_t\rangle$ can be evaluated by
\begin{equation}
\hat Q|\psi_t\rangle=\frac{i}{2\pi}\int^\infty_{-\infty}d\omega \hat G_{QP}(\omega+i0^+)e^{-i\omega t}|\psi_0\rangle,
\end{equation}
where $\hat G_{QP}(z)$ is another constrained propagator defined as
\begin{equation}
\hat G_{QP}(z)\equiv \hat Q(z-\hat H)^{-1}\hat P.
\end{equation}
One can check that 
\begin{equation}
\begin{split}
\hat G_P(z)&=\hat G^{(0)}_P(z)+\hat G^{(0)}_P(z)\hat H_{PQ}\hat G_{QP}(z) \\
\hat G_{QP}(z)&=\hat G^{(0)}_Q(z)\hat H_{QP}\hat G_P(z),
\end{split}
\label{GPGQP}
\end{equation}
where $\hat H_{PQ}\equiv \hat P\hat H\hat Q$, $\hat H_{QP}=\hat Q\hat H\hat P$ and
\begin{equation}
\begin{split}
\hat G^{(0)}_P(z)&\equiv\hat P(z -\hat H_P)^{-1}\hat P,\\
\hat G^{(0)}_Q(z)&\equiv\hat Q(z -\hat H_Q)^{-1}\hat Q,
\end{split}
\end{equation}
with $\hat H_P\equiv \hat P\hat H\hat P$ and $\hat H_Q\equiv \hat Q\hat H\hat Q$. Combining the two equations in Eq.~(\ref{GPGQP}) by eliminating $\hat G_{QP}(z)$, we obtain
\begin{equation}
\begin{split}
\hat G_P(z)&=[\hat G^{(0)}_P(z)^{-1}-\hat \Sigma(z)]^{-1},\\
\hat \Sigma(z)&=\hat H_{PQ}\hat G^{(0)}_Q(z)\hat H_{QP},
\end{split}
\label{GPz}
\end{equation}
which in turn implies [due to the second line in Eq.~(\ref{GPGQP})] that $\hat G_{QP}(z)$ can be expressed in terms of $\hat G^{(0)}_P(z)$, $\hat G^{(0)}_Q(z)$, $\hat H_{PQ}$ and $\hat H_{QP}$. 

While the above formulas are formally the same as the well-known results for Hermitian systems \cite{CohenTannoudji1998}, we emphasize that they apply equally to NH (lossy) systems and even NH projectors. As for the photon-emission problem discussed in the main text, however, the projector onto the atomic-excitation subspace is still Hermitian while both $\hat H_P$ and $\hat H_Q$ are NH in general. 
Concretely, we have
\begin{equation}
\begin{split}
&\hat H_P=\sum^N_{n=1}\Delta_n\hat\sigma^{ee}_n,\;\;\;\;\hat H_Q=\sum_{\boldsymbol{k}\in {\rm B.Z.}} \hat{\boldsymbol{a}}^\dag_{\boldsymbol{k}} h_{\boldsymbol{k}} \hat{\boldsymbol{a}}_{\boldsymbol{k}}, \\
&\hat H_{QP}=\hat H^\dag_{PQ} =\frac{1}{\sqrt{|\Lambda|}}\sum^N_{n=1}\sum_{\boldsymbol{k}\in{\rm B.Z.}}\hat\sigma^{ge}_n\hat{\boldsymbol{a}}^\dag_{\boldsymbol{k}} \boldsymbol{g}_{\boldsymbol{k}n},
\end{split}
\label{HPQQP}
\end{equation}
where all the notations follow those in Eq.~(\ref{eq:heff_momentum}). We focus on the single-excitation sector and choose the basis to be $\{\hat\sigma^{eg}_n|\boldsymbol{g}\rangle\}^N_{n=1}$ and $\{\hat a^\dag_{\boldsymbol{k}s}|{\rm vac}\rangle\}_{\boldsymbol{k}\in{\rm B.Z.},s\in I}$, which constitute the subspaces of the projectors $\hat P$ and $\hat Q$, respectively. By substituting Eq.~(\ref{HPQQP}) into Eqs.~(\ref{GPz}) and (\ref{GPGQP}), we obtain $G_P(z)=G_e(z)$ and $G_{QP}(z)=|\Lambda|^{-1/2}\bigoplus_{\boldsymbol{k}\in{\rm B.Z.}} G_{\boldsymbol{k}}(z)$, where $G_e(z)$ and $G_{\boldsymbol{k}}(z)$ are given in Eqs.~(\ref{Ge}) and (\ref{Gk}) in the main text, respectively.

\section{Vanishing self-energy at maximally winding regions \label{App:zero}}

We start detailing the calculation of the self-energy for the general Hatano-Nelson model,
\begin{align}
\begin{split}
    & \phi(z, x) \equiv \frac{1}{2\pi}\int_{-\pi}^\pi dk\frac{e^{ikx}}{z-h_k} \\
    & = \frac{1}{2\pi i} \oint dy \frac{y^{\abs{x}}}{a y^2 + b y + c} \\
    & = \frac{1}{2\pi i a(y_+ - y_-)}\oint dy \left(\frac{y^{\abs{x}}}{y - y_+} - \frac{y^{\abs{x}}}{y - y_-}\right) \\
    & = \frac{1}{\sqrt{\delta}}\left[y_+^{\abs{x}}\Theta(1 - \abs{y_+}) - y_-^{\abs{x}}\Theta(1 - \abs{y_-})\right], 
    \end{split}
    \label{eq:der_self_en_hn}
\end{align}
where we assume $a \neq 0$.
To go from the first line to the second we performed a change of variable $y=e^{i\sign(x)k}$, such that the integral becomes a contour integral along the unit circumference in the complex plane (counter-clockwise). The denominator of the integrand is a second-order polynomial in the new variable $y$ with coefficients: 
\begin{equation}
   a = \sign(x)\frac{\kappa}{2} - J\,, \ b = z + i\kappa\,, \ c = -\sign(x)\frac{\kappa}{2} - J \,. \label{eq:coefsabc}
\end{equation}
Then, we split the integral using partial fractions; $\delta = b^2 - 4ac =z^2+2i\kappa z - 4J^2$ denotes the discriminant of the second order polynomial, and $y_\pm =(-b \pm \sqrt{\delta})/(2a)$ denote its roots. 
Last, we use residue integration---only the poles inside the unit circle contribute to the integral, so we express the result with the Heaviside's step function $\Theta (y)$.
Eq.~(\ref{eq:der_self_en_hn}) is not valid when the denominator of the integrand is a first-order polynomial, which may happen if there is hopping only in one direction, e.g., when $J=\kappa/2$. In that case, we obtain the following result: For $x \geq 0$,
\begin{equation}
  \phi(z, x) = \begin{cases} 
    \frac{1}{z+i\kappa}
    \left(\frac{\kappa}{z + i\kappa}\right)^{x} \,, & \text{if } \abs{z+i\kappa} > \kappa,\\
   0 \,, & \text{if } \abs{z+i\kappa} < \kappa,
  \end{cases}
  \label{xp}
\end{equation}
whereas for $x < 0$
\begin{equation}
  \phi(z, x) = \begin{cases} 
    0 \,, & \text{if } \abs{z+i\kappa} > \kappa,\\
   -\frac{1}{\kappa}
      \left(\frac{\kappa}{z + i\kappa}\right)^{1 + x} \,, & \text{if } \abs{z+i\kappa} < \kappa.
  \end{cases}
  \label{xm}
\end{equation}

As we show next, some features of the self-energy are intimately related with the topology of the bath, as determined by the spectral winding number.
Let us remind the reader of the following theorem (known as the argument principle \cite{Ahlfors1979}): For any analytic function $f:\mathbb{C}\to\mathbb{C}$, the number of zeros counted with multiplicity inside the region delimited by the curve $\gamma: [a,b]
\to\mathbb{C}$, $\gamma(a) = \gamma(b)$, is equal to the winding number $\ind(f(\gamma))$ (cf. Eq.~(\ref{ind})) of the curve $f(\gamma)$ around the origin. Thus, for the polynomial $p(y) = ay^2 + by + c$, with coefficients specified by Eq.~\eqref{eq:coefsabc}, the number of roots inside the unit circle, which are the ones that contribute to the integral in Eq.~\eqref{eq:der_self_en_hn}, is the same as the index of the curve $p(e^{i k})$ for $k\in[-\pi,\pi]$. If $x\geq 0$, $p(e^{ik})=e^{ik}(z - h_k)$, which implies that $\ind(p(e^{ik})) = 1 + \ind(z - h_k)$. For $x\leq 0$, $p(e^{ik})=e^{ik}(z - h_{-k})$, thus, $\ind(p(e^{ik})) = 1 + \ind(z - h_{-k}) = 1 - \ind(z - h_k)$. Now we can clearly see that for points $z$ inside the loop defined by the bath's complex dispersion relation $h_k$, a situation that we will denote by $z\in 
\ell$, the self-energy strictly vanishes, $\phi(z\in \ell,x) = 0$, if $x\geq 0$ and $J > 0$, or if $x \leq 0$ and $J < 0$. 

In the following, we generalize this result to arbitrary 1D single-band NH lattices with finite range hopping. Denoting the largest leftward (rightward) hopping ranges as $q$ ($p$), the Bloch Hamiltonian (band dispersion) takes the following form:
\begin{equation}
h_k=\sum^q_{n=-p} J_n e^{ink},   
\label{hkpq}
\end{equation}
where the hopping amplitudes $J_n$'s are complex in general.

For the multi-emitter self-energy matrix, if the distance $x\ge 0$, the matrix element is proportional to the integral
\begin{equation}
\phi(z,x)=\oint_{|y|=1}\frac{dy}{2\pi i}\frac{y^{x+p-1}}{z y^p-\sum^{p+q}_{n=0} J_{n-p} y^n}.    
\label{pzx1}
\end{equation}
If $\ind(h_k - z)=-p\neq0$, which implies that no zeros of $z y^p-\sum^{p+q}_{n=0} J_{n-p} y^n$ are within the unit circle (according to the argument principle \cite{Ahlfors1979}), the above integral (\ref{pzx1}) vanishes. Similarly, if $x\le 0$, the matrix element is proportional to
\begin{equation}
\phi(z,x)=\oint_{|y|=1}\frac{dy}{2\pi i}\frac{y^{-x+q-1}}{z y^q-\sum^{p+q}_{n=0} J_{-n+q} y^n}.    
\end{equation}
If $\ind(h_k - z)=q\neq0$, which implies that no zeros of $z y^q-\sum^{p+q}_{n=0} J_{-n+q} y^n$ are within the unit circle, the above integral vanishes. 

Note that $\{z:\ind(h_k-z)=-p\}$ or $\{z:\ind(h_k-z)=q\}$ is the maximally winding regions, since the spectral winding number of $h_k$ in Eq.~(\ref{hkpq}) can only take on integers among $[-p,q]$. For the Hatano-Nelson model ($p=1$ and/or $q=1$), the maximally winding region is simply the interior of the spectral loop.

\begin{figure}
    \centering
    \includegraphics[width=8cm, clip]{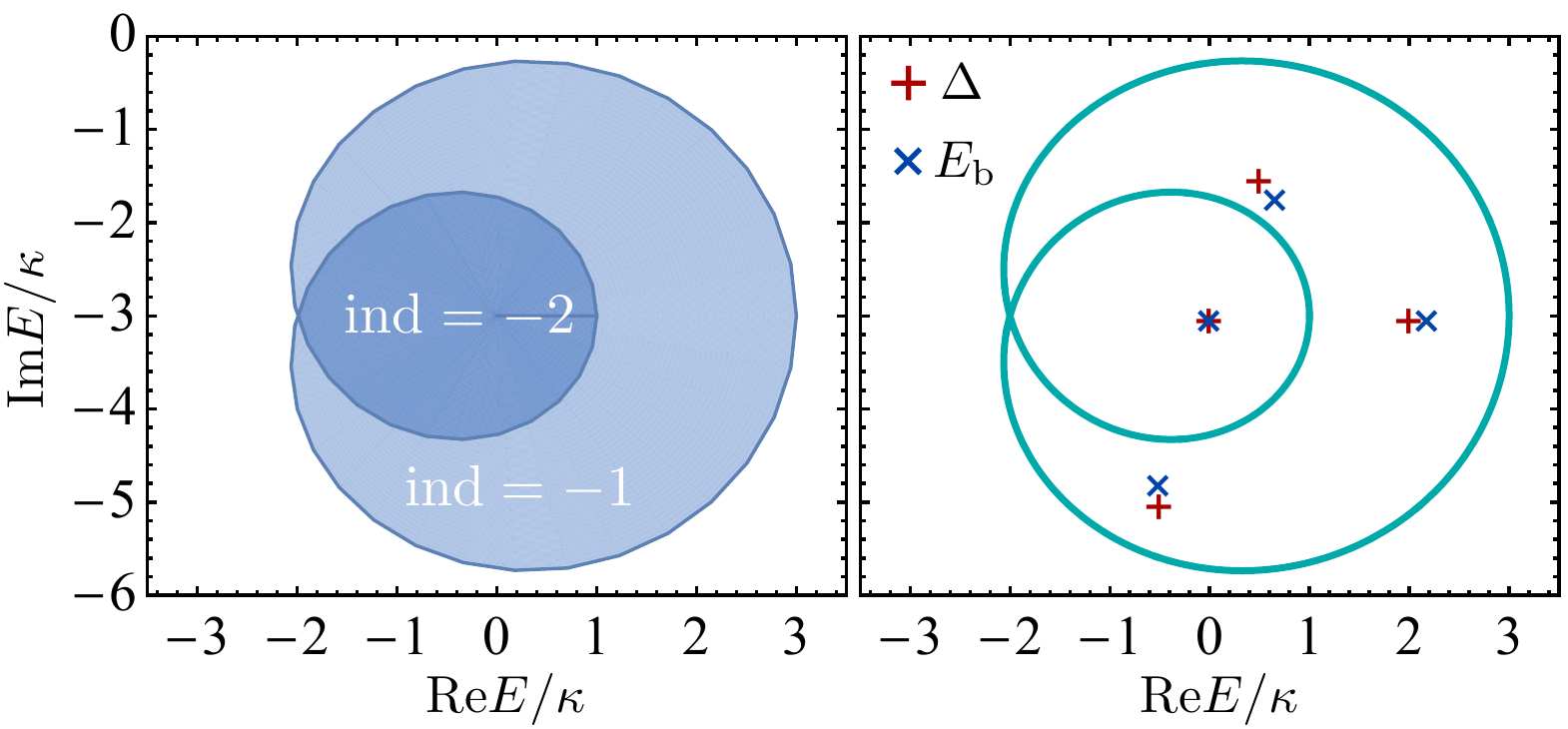}
    \caption{For the NH lattice described by Eq.~(\ref{hk12}) with $\kappa'=2\kappa$, there are both maximally ($\ind =-2$) and non-maximally ($\ind = -1$) winding regions in the complex energy plane (left). While the eigenenergy $E_{\rm b}$ of the hidden bound state in the maximally winding region is always pinned at the complex detuning $\Delta$, such a relation generally breaks down in the non-maximally winding region (right).}
    \label{fig:nonmax}
\end{figure}

Let us provide another simple example showing that a nonzero winding number alone (i.e., not maximal in general) is not sufficient for pinning the eigenenergy of the (hidden) bound state at the complex detuning. Consider a 1D NH lattice with both NN and next-to-NN unidirectional right hopping. The Bloch Hamiltonian reads
\begin{equation}
 h_k = \kappa (e^{-ik} - 1) + \kappa' (e^{-2ik} - 1).  
 \label{hk12}
\end{equation}
The corresponding self-energy can be analytically obtained to be
\begin{equation}
\Sigma(z)= \frac{g^2}{\sqrt{\delta}}(y_+\Theta_+ - y_-\Theta_-),
\label{SENNNN}
\end{equation}
where $y_\pm$ is the roots of $(z+i\kappa+i\kappa')y^2 - \kappa y - \kappa'=0$, $\delta =\kappa^2 + 4\kappa'(z+i\kappa+i\kappa')$ is the discriminant and $\Theta_\pm\equiv \Theta(1-|y_\pm|)$. By choosing, e.g., $\kappa'=2\kappa$, we find that there is a non-maximally winding region ($\ind=-1$) on top of a maximally winding region ($\ind = -2$), as shown in Fig.~\ref{fig:nonmax}. One can check that the self-energy (\ref{SENNNN}) identically vanishes in the maximally winding region, but otherwise has a nontrivial $z$ dependence. As a result, if a complex detuning $\Delta$ lies in a non-maximally winding region, we do not have a bound state with eigenenergy being exactly $\Delta$ in general (cf. Fig.~\ref{fig:nonmax}).

\section{Branch-cut singularities for the alternatingly lossy lattice (\ref{1DPT})}
\label{App:PTBC}
We provide quantitative explanations on the various exponents of algebraic atom decay observed in the passive $PT$-symmetric lattice (\ref{1DPT}) with alternating loss. Since we are only interested in $|c_e(t)|^2$, it suffices to know the absolute value of $F$. Accordingly, we do not need to exactly identify $\Sigma_{\rm l/r}$ but only have to know the amount of sudden jump in the self-energy. Noting that $y_+y_-=1$ for $y_\pm$ in Eq.~(\ref{PTSE}), we know that the self-energy right before/after the jump is nothing but the coefficient before $\Theta_+$/$\Theta_-$ (or vice versa).

For a single-emitter in sublattice $A$, the self-energy is given by Eq.~(\ref{SA}). Therefore, the lhs of Eq.~(\ref{Fnu}) (up to a sign, as is always the case in this Appendix) reads
\begin{equation}
\frac{1}{z-\Delta - \frac{g^2z}{\sqrt{\delta}}} - \frac{1}{z-\Delta + \frac{g^2z}{\sqrt{\delta}}}= \frac{2g^2z\sqrt{\delta}}{\delta(z-\Delta)^2 - g^4 z^2},
\label{SEA}
\end{equation}
where $\delta=z(z+i\kappa)(z^2 + i\kappa z - 4J^2)\simeq -4i\kappa J^2 z$ for small $z$ (i.e., $z$ close to $0$, which is the branch point). If $\Delta=0$, the denominator of the rhs of Eq.~(\ref{SEA}) is dominated by $-g^4z^2$ for small $z$, leading to $\nu=-1/2$ and $|F|=4J\sqrt{\kappa}/g^2$. Otherwise, the denominator is dominated by $\delta\Delta^2$, leading to $\nu=1/2$ and $|F|=g^2/(J\sqrt{\kappa}|\Delta|^2)$.

Similarly, for a single-emitter in sublattice $B$, the self-energy is given by Eq.~(\ref{SB}). The lhs of Eq.~(\ref{Fnu}) thus reads
\begin{equation}
\begin{split}
&\frac{1}{z-\Delta - \frac{g^2(z+i\kappa)}{\sqrt{\delta}}} - \frac{1}{z-\Delta + \frac{g^2(z+i\kappa)}{\sqrt{\delta}}} \\
=& \frac{2g^2(z+i\kappa)\sqrt{\delta}}{\delta(z-\Delta)^2 - g^4 (z+i\kappa)^2},
\end{split}
\label{SEB}
\end{equation}
where $\delta$ follows that in Eq.~(\ref{SEA}) (same applies hereafter). Unlike the previous case, now the denominator of the rhs of Eq.~(\ref{SEB}) is always dominated by $g^4\kappa^2$ for small $z$ regardless of $\Delta$, leading to $\nu=1/2$ and $|F|=4J/(g^2\sqrt{\kappa})$. The results so far have already been mentioned in the main text.

We move on to the case of two emitters located in sublattice $A$. The initial state is assumed to be in a symmetric superposition, so the self-energy reads
\begin{equation}
\Sigma(z) = \Sigma^{AA}_{11}(z) + \Sigma^{AA}_{12}(z),
\end{equation}
where $\Sigma^{AA}_{mn}(z)$ is given in Eq.~(\ref{PTSE}). If the two emitters are in the NN unit cells, we have $|x_{12}|=1$ in Eq.~(\ref{PTSE}) and the lhs of Eq.~(\ref{Fnu}) turns out to be
\begin{equation}
\begin{split}
&\frac{1}{z-\Delta - \frac{g^2 z}{\sqrt{\delta}}(1+y_+)} - \frac{1}{z-\Delta + \frac{g^2 z}{\sqrt{\delta}}(1+y_-)} \\
=&\frac{g^2\sqrt{\delta} z^2(z+i\kappa)}{J^2\delta(z-\Delta)^2-g^2\delta z(z-\Delta)-g^4z^3(z+i\kappa)}.
\end{split}
\label{2EA1}
\end{equation}
The numerator of the rhs of Eq.~(\ref{2EA1}) is always of $\mathcal{O}(z^{5/2})$ for small $z$. If $\Delta=0$, the denominator is of $\mathcal{O}(z^3)$, leading to $\nu=-1/2$. However, the corresponding $t^{-1}$ decay is invisible due to the (accidentally) stable bound states. Otherwise, the denominator is of $\mathcal{O}(z)$, leading to $\nu=3/2$. The corresponding $t^{-5}$ decay is visible (at least in the long-time limit) since now the bound state has a nonzero imaginary energy.

If the emitters are located in two next-to-NN unit cells, we have $|x_{12}|=2$ in Eq.~(\ref{PTSE}) and the lhs of Eq.~(\ref{Fnu}) turns out to be
\begin{equation}
\begin{split}
&\frac{1}{z-\Delta - \frac{g^2 z}{\sqrt{\delta}}(1+y^2_+)} - \frac{1}{z-\Delta + \frac{g^2 z}{\sqrt{\delta}}(1+y^2_-)} \\
=&\frac{g^2b^2\sqrt{\delta} z }{J^4\delta(z-\Delta)^2-g^2b\delta z(z-\Delta)-g^4b^2z^2},
\end{split}
\label{2EA2}
\end{equation}
where $b=-2J^2+z(z+i\kappa)\simeq -2J^2$ for small $z$. The numerator of the rhs of Eq.~(\ref{2EA2}) is always of $\mathcal{O}(z^{3/2})$. If $\Delta=0$, the denominator is of $\mathcal{O}(z^2)$, leading to $\nu=-1/2$ and a corresponding $t^{-1}$ decay. Otherwise, the denominator is of $\mathcal{O}(z^2)$, leading to $\nu=1/2$ and a corresponding $t^{-3}$ decay. All of these results have been numerically demonstrated in Fig.~\ref{fig:two_emitter_algebraic}(a).

\bibliography{GZP_references,MBG_references,library}

\end{document}